\documentclass[aps,prd,twocolumn,showpacs,nofootinbib,amsmath,amssymb,floatfix,superscriptaddress,showkeys]{revtex4}
\usepackage{graphicx}% Include figure files
\usepackage{dcolumn}% Align table columns on decimal point
\usepackage{bm}% bold math
\usepackage{captcont}
\usepackage{xcolor}

\usepackage{epstopdf}
\usepackage{mathtools}
\usepackage{natbib}
\usepackage{ulem}
\definecolor{blue0}{rgb}{0,0,0.6}
\usepackage[colorlinks,linkcolor=blue0,anchorcolor=blue0,citecolor=blue0,urlcolor=blue0]{hyperref}

\newcommand{\jcap}{J. Cosmol. Astropart. Phys.}
\newcommand{\apjl}{Astrophys. J. Lett.}

\newcommand{\mnras}{Mon. Not. R. Astron. Soc.}

\begin{document}
\title{Search for line-like signals in the all-sky Fermi-LAT data}
\author{Shang Li}
\affiliation{Key Laboratory of Dark Matter and Space Astronomy, Purple Mountain Observatory, Chinese Academy of Sciences, Nanjing 210008, China}
\affiliation{University of Chinese Academy of Sciences, Yuquan Road 19, Beijing, 100049, China}
\author{Zi-Qing Xia}
\affiliation{Key Laboratory of Dark Matter and Space Astronomy, Purple Mountain Observatory, Chinese Academy of Sciences, Nanjing 210008, China}
\affiliation{School of Astronomy and Space Science, University of Science and Technology of China, Hefei, 230026, China}
\author{Yun-Feng Liang}
\email{liangyf@pmo.ac.cn}
\affiliation{Key Laboratory of Dark Matter and Space Astronomy, Purple Mountain Observatory, Chinese Academy of Sciences, Nanjing 210008, China}
\author{Kai-Kai Duan}
\affiliation{Key Laboratory of Dark Matter and Space Astronomy, Purple Mountain Observatory, Chinese Academy of Sciences, Nanjing 210008, China}
\affiliation{University of Chinese Academy of Sciences, Yuquan Road 19, Beijing, 100049, China}
\author{Zhao-Qiang Shen}
\affiliation{Key Laboratory of Dark Matter and Space Astronomy, Purple Mountain Observatory, Chinese Academy of Sciences, Nanjing 210008, China}
\affiliation{University of Chinese Academy of Sciences, Yuquan Road 19, Beijing, 100049, China}
\author{Xiang Li}
\affiliation{Key Laboratory of Dark Matter and Space Astronomy, Purple Mountain Observatory, Chinese Academy of Sciences, Nanjing 210008, China}
\author{Lei Feng}
\affiliation{Key Laboratory of Dark Matter and Space Astronomy, Purple Mountain Observatory, Chinese Academy of Sciences, Nanjing 210008, China}
\author{Qiang Yuan}
\affiliation{Key Laboratory of Dark Matter and Space Astronomy, Purple Mountain Observatory, Chinese Academy of Sciences, Nanjing 210008, China}
\affiliation{School of Astronomy and Space Science, University of Science and Technology of China, Hefei, 230026, China}
\author{Yi-Zhong Fan}
\email{yzfan@pmo.ac.cn}
\affiliation{Key Laboratory of Dark Matter and Space Astronomy, Purple Mountain Observatory, Chinese Academy of Sciences, Nanjing 210008, China}
\affiliation{School of Astronomy and Space Science, University of Science and Technology of China, Hefei, 230026, China}
\author{Jin Chang}
\affiliation{Key Laboratory of Dark Matter and Space Astronomy, Purple Mountain Observatory, Chinese Academy of Sciences, Nanjing 210008, China}
\date{\today}

\begin{abstract}
In order to search for the line-like signals in the Fermi-LAT data, we have analyzed totally 49152 regions of interest (ROIs) that cover the whole sky. No ROI displays a line signal with test statistic (TS) value above 25, while for 50 ROIs weak line-like excesses with ${\rm TS}>16$ are presented. The intrinsic significances of these potential signals are further reduced by the large trial factor introduced in such kind of analysis. For the largest TS value of 24.3 derived in our analysis, the corresponding global significance is only $0.54\sigma$. We thus do not find any significant line-like signal and set up constraints on the cross section of dark matter annihilating to gamma-ray lines, $\left<\sigma{v}\right>_{\gamma\gamma}$.
\end{abstract}

\maketitle
\section{Introduction}

A gamma-ray line signal, if robustly detected, will be deemed as a smoking-gun signature of particle dark matter (DM) since no known astrophysical process can generate such a specific spectrum, while the line signal in principle can be from the direct annihilation of DM particles into gamma-rays (i.e., $\chi\chi\rightarrow\gamma\gamma, \gamma{Z}$ or $\gamma{H}$). It may be captured if the annihilation cross section is large enough that the line signal exceeds the detection sensitivities of gamma-ray telescopes such as Fermi-LAT \cite{atwood09LAT} and DAMPE \cite{dampe}. Great efforts have been made to hunt for such kind of signal, but none is conclusively detected so far \cite{pullen07EGRETline,fermi10line1,fermi12line2,bringmann12_130gev,weniger12_130gev,geringer12dsphLine,tempel12_130gev,huang12_130gev,su2012line,fermi13line,Hektor2013gcline,albert14line,fermi15line,anderson16gclsLine,liang16gclsLine,Profumo2016line,liang16dsphLine,Liang17line3}.

Some tentative evidence of line signals has been suggested in the literature. By analyzing an optimized region around the Galactic center, a line-like excess at 130 GeV was found in the 4 years' Fermi-LAT Pass 7 data \cite{bringmann12_130gev, weniger12_130gev}. 
This signal was also reported with lower significance in the searches of galaxy clusters \cite{Hektor2013gcline}. However, analysis by Fermi-LAT collaboration using 5.8 years of Pass 8 data did not confirm this signal \cite{fermi15line}. More recently, a tentative line-like excess at 42.7 GeV was found in the stacked spectrum of 16 nearby galaxy clusters, the global significance of this excess is just $\sim3.0\,\sigma$ \cite{liang16gclsLine,fl16}.

The most promising site one might be able to observe a line signal is the region around the center of our Milky Way. Besides the Galactic center (GC), other regions that may produce considerable line signals include dwarf spheroidal galaxies (dSphs) \cite{geringer12dsphLine,liang16dsphLine}, DM subhalos \cite{tempel12_130gev,Liang17line3} and galaxy clusters \cite{Hektor2013gcline,anderson16gclsLine,liang16gclsLine}. 
In this work, we do not examine specific objects/regions, but perform blind searches for the line signals in the whole sky using the Fermi-LAT data. 
We aim to find out some regions with relatively high TS values. Due to the very large trial factor introduced in such kind of analysis, none of the weak excesses can not be identified as a real signal. Nevertheless, our search results may be taken as a list of regions that is worth further attention, since the possibility that very a few of them are DM line signals from subhalos can not be ruled out.
We also set limits on the DM properties utilizing the number of the line-like excesses (see Sec. \ref{sec:limits}).

\section{Fermi-LAT data and line signal search}
\label{sec2}
In this work, we use the Fermi-LAT data to perform the searches\footnote{\url{https://fermi.gsfc.nasa.gov/ssc/data/}}. We will search for the signals with line energies from 5 GeV to 300 GeV, thus we take into account the Fermi-LAT data in the energy range of 1 GeV to 500 GeV to address the energy dispersion of the instrument.
The time period of the data we use is from Aug. 4th, 2008 to Aug. 4th, 2017 (corresponding to MET 239557417-523497605). We take the recommended zenith angle cut ($\theta_{\rm zenith}<90^{\circ}$) and data quality cut ({\tt DATA\_QUAL==1 \&\& LAT\_CONFIG==1}) to avoid the contamination from Earth limb emission and to guarantee the data is suitable for science use.
%\footnote{\url{https://fermi.gsfc.nasa.gov/ssc/data/analysis/documentation/Cicerone/Cicerone_Data_Exploration/Data_preparation.html}}. 
To reduce the contamination from residual cosmic rays in the LAT data, and also to be consistent with our previous works \cite{liang16gclsLine,liang16dsphLine,Liang17line3}, we make use of the {\tt ULTRACLEAN} data. For achieving better energy resolution, we exclude the {\tt EDISP0} data in our analysis (${\tt evtype=896}$). We use the {\tt Fermi Science Tools} of version {\tt v10r0p5} to do the data selection and the exposure calculation.

%\begin{table*}[!htb]
%\caption{9 ROIs give the highest TS values of the line signals. }
%\begin{ruledtabular}
%\begin{tabular}{ccccccc}
% \#\footnote{The {\tt HEALPix} index in {\tt NESTED} ordering.} &  Energy [GeV] & TS Value & LON [$^\circ$] & LAT [$^\circ$] & RA [$^\circ$] & Dec [$^%\circ$]\\
% \hline
%25363 & 74.9 & 24.3 & 182.81 & -15.09 &  74.40 &  18.21 \\
%21714 & 17.5 & 22.4 & 114.61 &  -5.98 & 357.97 &  55.93 \\
% 9695 & 29.2 & 21.5 & 269.15 &  50.48 & 171.98 &  -6.83 \\
%45226 & 16.2 & 21.0 & 272.81 & -78.28 &  20.20 & -37.07 \\
%15172 & 48.6 & 20.1 & 296.32 &  50.48 & 188.56 & -12.17 \\
%24414 &  9.4 & 20.0 &  97.73 &  31.39 & 265.83 &  67.66 \\
%34394 & 18.2 & 19.6 &  62.58 & -41.81 & 332.47 &   1.38 \\
%18272 & 38.4 & 19.6 &  25.31 &   7.78 & 272.45 &  -3.16 \\
%37762 &  5.2 & 19.5 & 125.36 & -58.92 &  14.11 &   3.93 \\
%\end{tabular}
%\end{ruledtabular}
%\label{tb1}
%\end{table*}

To search for the line signals in the whole sky, we select totally 49152 ROIs with a radius of 2 degree for each. The centers of the ROIs correspond to the {\tt HEALPix} \cite{healpix05} coordinates list with {$\tt nside = 64$}. Such a strategy ensures that all the sky is covered by our ROI sample. Assuming a point-like spatial distribution for the line signal, the $2^\circ$ radius also ensures that most line signal photons are included by the ROI even the signal is located at the edge of a {\tt HEALPix} pixel considering the point spread function (PSF) of Fermi-LAT is smaller than $1^\circ$ for $>5\,{\rm GeV}$ data \cite{fermi12cal} and the radius of the pixel is roughly 0.5 degree\footnote{The shape of the {\tt HEALPix} pixel is in fact not a circle, the radius here is an estimation derived using the solid angle of each pixel.}.

In each ROI, the sliding window technique \cite{pullen07EGRETline,weniger12_130gev,liang16gclsLine} is adopted to perform the search. 
For each putative line with energy $E_{\gamma}$, we perform unbinned likelihood fittings in a narrow window of ($E_{\gamma}-0.5E_{\gamma}$, $E_{\gamma}+0.5E_{\gamma}$). The test statistics (TS) is obtained by comparing the likelihoods of null model (no line signal model) and the signal model. We approximate the null model to a power law function. In consideration of that the background mixing all astrophysical components should be smooth and continuous in spectra, the power law approximation is reasonable since we are using a very narrow energy window. For the signal model, we adopt the form of a line component ($\delta(E-E_\gamma)$) superposing on the power law background. 
For the line component, we have also convolved it with the energy dispersion function of the data. 
The method of searching for line signals in the Fermi-LAT data has been extensively introduced in Ref.\cite{weniger12_130gev,fermi13line,fermi15line,liang16gclsLine}. We refer readers to these literature for details.

\section{results}
\begin{figure}[!htb]
\includegraphics[width=0.9\columnwidth]{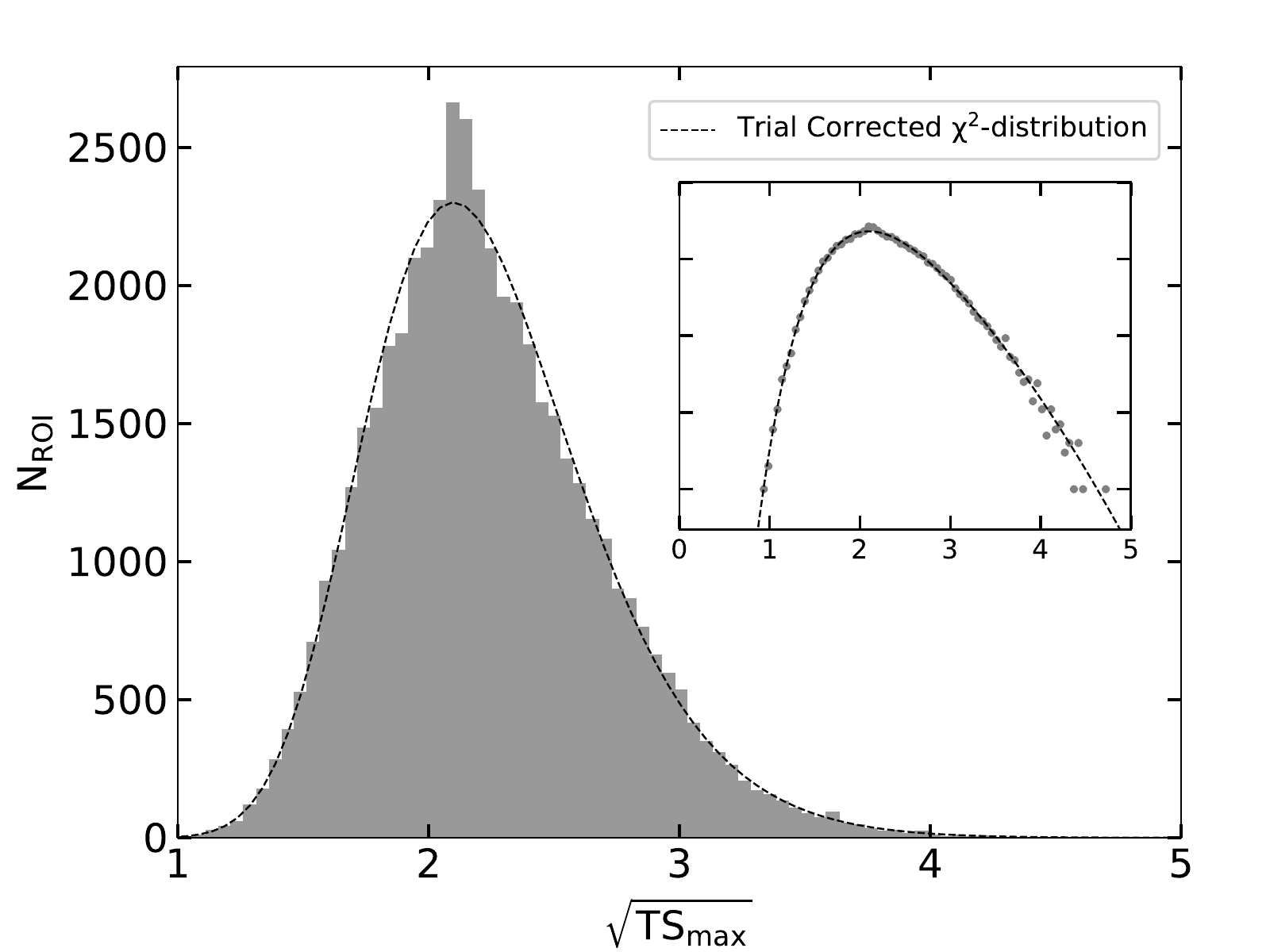}
\caption{The $TS_{\rm max}$ distribution of 49152 searched ROIs. The dashed line is the best fit trial-corrected $\chi^2$ distribution. The inserted sub-figure is the same distribution plotted in a y-log framework for better showing the tails of the distribution.}
\label{fig:tsdistribution}
\end{figure}

Adopting the aforementioned approach, we searched totally 49152 ROIs for the line signals.  We summarize our search results in this section.
Figure \ref{fig:tsdistribution} presents the $TS_{\rm max}$ distribution over all the ROIs. The $TS_{\rm max}$ denotes the maximum TS value among a series of attempted line energies\footnote{Explicitly, 110 $E_\gamma$ in the range of 5$-$300$\,{\rm GeV}$.} in each ROI. As expected, most of ROIs give relatively low TS values ($TS_{\rm max}<9$ for 94\% ROIs). Theoretically, the $TS_{\rm max}$ for the background only data should follow a trial-corrected $\chi^2$ distribution \cite{weniger12_130gev,liang16gclsLine}.  
%We find that this distribution can fit our search results well (dashed line in Figure \ref{fig:tsdistribution}).
Fitting our results with this distribution gives $\chi^2_{\rm red}=202.5/58$, indicating the best fit can not match the data well. Considerable discrepancy between the best fit curve and the TS distribution is clearly seen around $TS_{\rm max}\sim4-5$. To check whether the deviation is artificially from the analysis method we use, we have made some tests in Appendix \ref{app:test}. Using the same fitting code to analyze MC simulation data, we obtain results well consistent with the theoretical predictions. We therefore conclude that the deviation is not related to our fitting method. It may come from the non-poisson background of the real events due to systematics related to instrument
measurements or induced by the approximation of a power law background in each energy window \cite{fermi15line}. 
However, we find the tails of the curve can match the histogram relatively well, it is still reasonable to use the best-fit function to approximate the null distribution (i.e., the distribution for background-only data) for large $TS_{\rm max}$.

In our all-sky searches, no signal is found to have a TS value greater than 25 ($TS=25$ corresponds to a local significance of $5\,\sigma$).
The most significant line-like excess appears in the ROI centered on ($l=182.81$, $b=-15.09$), with a TS value of 24.3. The corresponding line energy is 74.9 GeV. 
%We plot the observation spectrum of this ROI in the first panel of Figure \ref{fig:spectrum}. 
The observation spectrum of this ROI can be found in Appendix \ref{spectrum}.
Evidence of excess is clearly seen for this ROI, indicating that our search strategy can effectively identify such a kind of signal in the spectrum. 
Besides this excess, in total 50 ROIs give $TS_{\rm max}>16$ and 2953 result in $TS_{\rm max}>9$.
 
%The 9 ROIs give the highest TS values of line signal are listed in Table \ref{tb1} 
%We also show the spectra of some selected ROIs in Figure \ref{fig:spectrum}. 

\begin{figure*}[!t]
\includegraphics[width=0.9\textwidth]{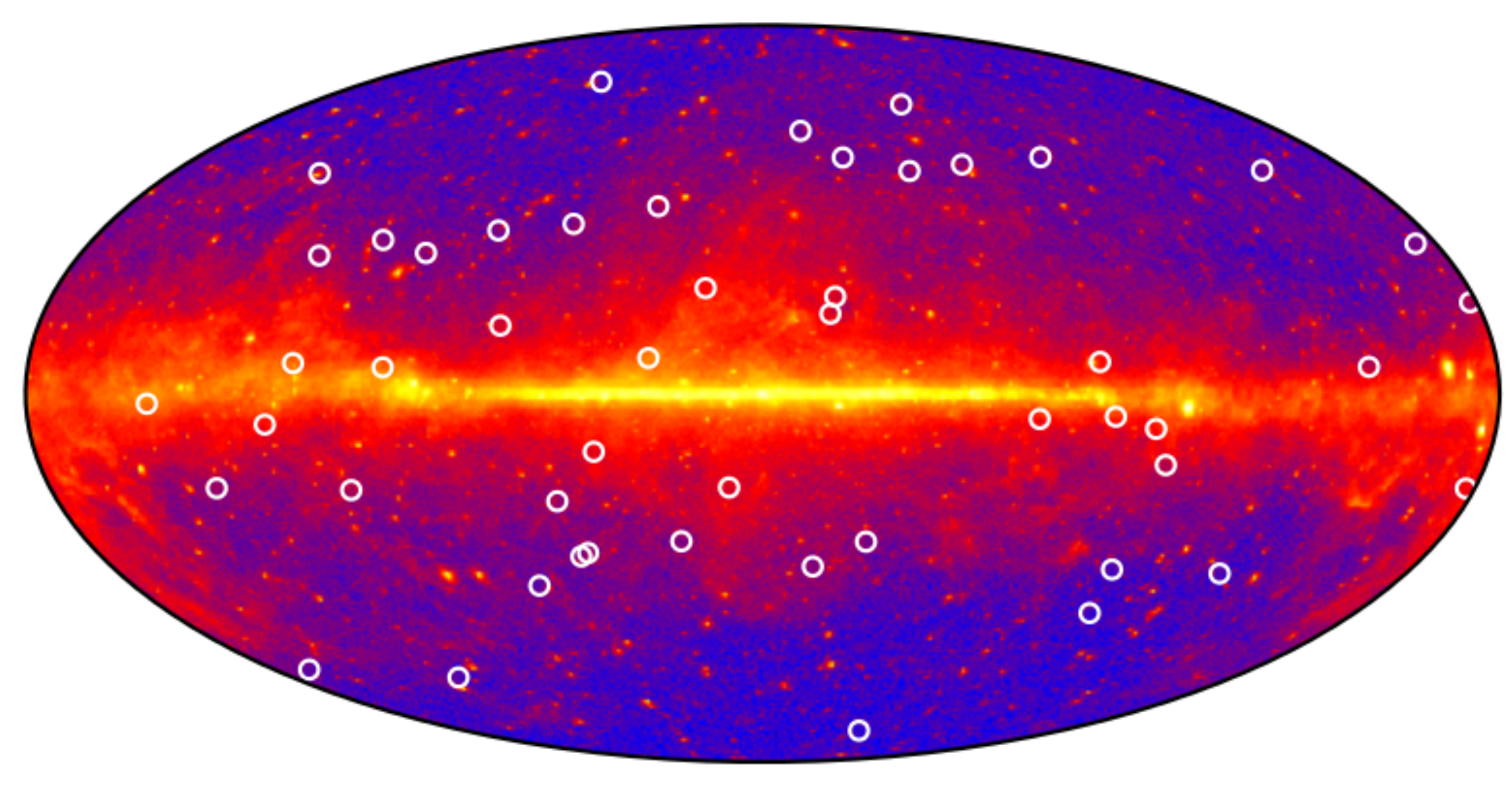}
\caption{ROIs showing line-like excesses with $\rm TS >16$ overlaid on a Hammer-Aitoff projection of Fermi-LAT counts map ($E>1\,{\rm GeV}$).}
\label{fig:dis}
\end{figure*}

The 50 ROIs showing highest TS values are of particular interest to us, because they have higher probabilities of being from real signal rather than background fluctuation.
We plot their positions in the sky in Figure \ref{fig:dis}.
Though these regions have $TS_{\rm max}>16$, intrinsically their global significances are very low.
The reason is that we have searched a lot of ROIs and for each ROI a series of line energies, an extremly large trial-factor is introduced if we convert the TS values to the global significances. 
%Thus the {\bf line-like excesses} presented here have intrinsically low significance. 
Utilizing the null distribution in Figure \ref{fig:tsdistribution} and attributing 49152 trials to the scan over multiple ROIs, the derived global significances are 0.54$\sigma$ and 0.11$\sigma$ for the first two ROIs with $TS_{\rm max}=24.3$ and $22.4$, respectively, and $<0.1\sigma$ for any other ROIs.
These weak line-like excesses are most likely from statistical fluctuations. In view of the great importance of the line signal, these regions still deserve further attention. 
If one can find the same excesses in some of these regions by analyzing the data from other gamma-ray telescopes,
the statistical origin will be disfavored. 
%Table I offers a list of {\bf potential line signal regions} for later studies.
%All the ROIs with $TS_{\rm max}>16$ have been presented in Table \ref{tb:50roi}.
We thus present the information of the ROIs with $TS_{\rm max}>16$ in Appendix \ref{list}, they can be treated as a list of potential line signal regions for later studies.

\section{Searching for counterparts of the weak line-like excesses}
\label{sec4}
In addition to waiting for the observations from other instruments, we have also attempted to test the possible DM origin of these excesses by searching for their counterparts.
The dark matter particles that generate the gamma-ray lines may simultaneously annihilate to other Standard Model particles (e.g., $b\bar{b}$, $\tau^+\tau^-$) \cite{Lefranc:2016fgn}, which could yield continuum gamma-ray emission at lower energies. This model-independent gamma-ray emissions provide us a way to test the possible DM origin of the line-like excesses.
Specifically, for a certain excess, if we could detect another gamma-ray component in the ROI with its spectrum and spatial distribution compatible with DM continuum emission, it is a strong evidence that both the line and continuum emission are from DM annihilation within a given subhalo. For this reason, we analyze the unassociated point sources within the ROIs of $TS_{\rm max}>16$. Due to the complicated gamma-ray backgrounds in the Galactic plane, we ignore the ROIs with latitudes $|b|<10^\circ$. Totally, 13 unassociated point sources in FL8Y\footnote{\url{https://fermi.gsfc.nasa.gov/ssc/data/access/lat/fl8y/}} are found. We apply the standard likelihood analysis of Fermi-LAT data
%\footnote{\url{https://fermi.gsfc.nasa.gov/ssc/data/analysis/scitools/binned_likelihood_tutorial.html}} 
to these unassociated sources in the energy range from 300 MeV to 300 GeV. The unassociated point sources are modeled with spectrum of DM annihilation\footnote{The DM spectra are implemented with DMFitFunction: \url{https://fermi.gsfc.nasa.gov/ssc/data/analysis/scitools/source_models.html}} and the spectral function adopted in FL8Y. For the DM model, we consider the annihilation channel of $b\bar{b}$ and $\tau^+\tau^-$, and the DM mass is fixed to the $E_\gamma$ giving the largest TS value in each ROI. The delta likelihood between these two models, $\Delta\ln{\mathcal L}=\ln{{\mathcal L}_{\rm DM}}-\ln{{\mathcal L}_{\rm FL8Y}}$, is used to determine whether a DM annihilation hypothesis is favored. 
%The $\Delta\ln{\mathcal{L}}$ for the 13 unassociated sources are listed in Table \ref{tb:13src}. 
Our analyses show that only 1 among 13 sources, FL8Y J1656.4-0410, marginally favors the DM spectrum over the spectral model used in FL8Y. The $\Delta\ln{\mathcal L}=7.0$ corresponds to a local significance of $<4\sigma$, not offering an evidence of DM signal from subhalo.

%%############################################################
%\begin{table}[!htb]
%\caption{The 13 unassociated point sources within selected ROIs.}
%\begin{ruledtabular}
%\begin{tabular}{cccccccc}
% Name & {$\Delta \ln{\mathcal L}$} & RA [$^\circ$] & Dec [$^\circ$] &  \#\footnote{The ROIs where the sources are located.}\\
% \hline
% FL8Y J0459.4$+$1921 & -4.35    &  74.87 &  19.36  & 25363 \\
% FL8Y J0500.6$+$1904 & -3.5     &  75.16 &  19.07  & 25363 \\
% FL8Y J1238.5$-$1159 & -48.8    & 189.65 & -11.99  & 15172 \\
% FL8Y J1947.0$+$0030 & -7.7    & 296.75 &   0.52  & 36744 \\
% FL8Y J1959.3$-$6121 & -8.7     & 299.84 & -61.36  & 46981 \\
% FL8Y J0741.0$-$5226 & -7.6     & 115.27 & -52.44  & 29481 \\
% FL8Y J1555.6$-$2907 & -0.9     & 238.92 & -29.13  & 20103 \\
% FL8Y J1405.2$-$0641 & -8.6     & 211.32 &  -6.70  & 14243 \\
% FL8Y J1105.9$+$7304 & -2.4     & 166.48 &  73.07  & 7168  \\
% FL8Y J1100.7$+$7018 & -4.5     & 165.19 &  70.31  & 7168  \\
% FL8Y J1656.4$-$0410 & 7.0      & 254.12 &  -4.17  & 19922 \\
% FL8Y J1612.0$+$1407 & -33.1    & 243.01 &  14.12  & 2458  \\
% FL8Y J1544.2$-$2554 & -30.0  & 236.06 & -25.91  & 20147 \\
%\end{tabular}
%\end{ruledtabular}
%\label{tb:13src}
%\end{table}

Besides, if certain of the line-like excesses in the Table \ref{tb:50roi} is a real DM signal, it could be from any channel of $\chi\chi\rightarrow\gamma\gamma$ or $\gamma{Z}$ or $\gamma{H}$, then it is possible that the dark matter particles also annihilate through another channel among the three \cite{su2012line}. For example, assuming the first line signal is from $\chi\chi\rightarrow\gamma\gamma$, the second line would be located at $E'_\gamma=m_\chi(1-m_X^2/4m_\chi^2)$, where $X$ could be $Z$ or $h$. 
If the second line is found with high significance, it also offers an indication that the line-like excess in Table \ref{tb:50roi} is a real DM signal. Thus for the 50 ROIs, we calculate the TS values of the second line signals at the corresponding energies. The largest TS values for the second lines are listed in Table \ref{tb:50roi} as well. We find that only 4 ROIs result in TS$_{\rm 2nd}>4$, and the highest one appears in the ROI {\tt \#35888}. The combined TS of this ROI reaches 27.1 for the two gamma-ray lines, however considering additional degree of freedom and very large trial factors, the global significance is still very low.

\section{Constraining DM cross section with the non-detection of significant line signal}
\label{sec:limits}
In the cold dark matter paradigm, structure forms hierarchically, and it is predicted that there exist large amount of DM subhalos around the Milky Way. Such a prediction is supported by numerical $N$-body simulations \cite{diemand08VL2,springel08Aquarius,garrison14ELVIS}. The concentration of DM in the subhalos leads to a higher DM annihilation rate. If massive subhalos are close to the Earth sufficiently, they may generate gamma-ray signals detectable by Fermi-LAT.
For some subhalos which are too small to capture enough baryonic matter (i.e., $M_{\rm sub}<10^{8}\,M_{\odot}$), the gamma-ray annihilation signals would be the only channel to observe them. Thus, it is supposed that some unassociated Fermi-LAT sources are potential DM subhalos \cite{fermi12dmsh,berlin14,bertoni15dmsh,schoonenberg16dmsh,bertoni16j2212,hooper16dmsh,calore16dmsh,wyp16dmsh,xzq16dmsh}, especially those spatially extended and with spectra compatible with DM signals \cite{bertoni16j2212,wyp16dmsh,xzq16dmsh}.

\begin{figure*}[!htb]
\includegraphics[width=0.45\textwidth]{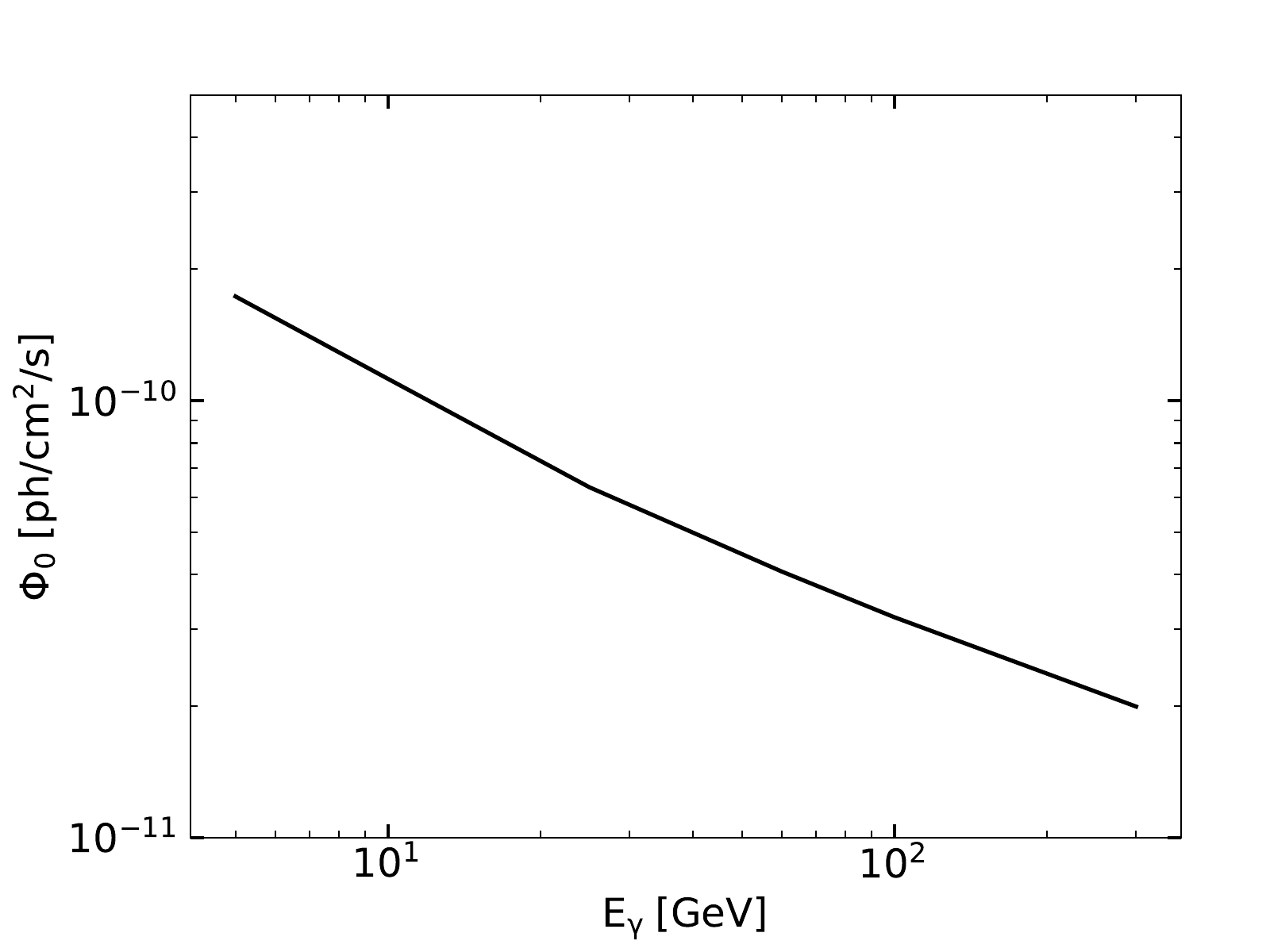}
\includegraphics[width=0.45\textwidth]{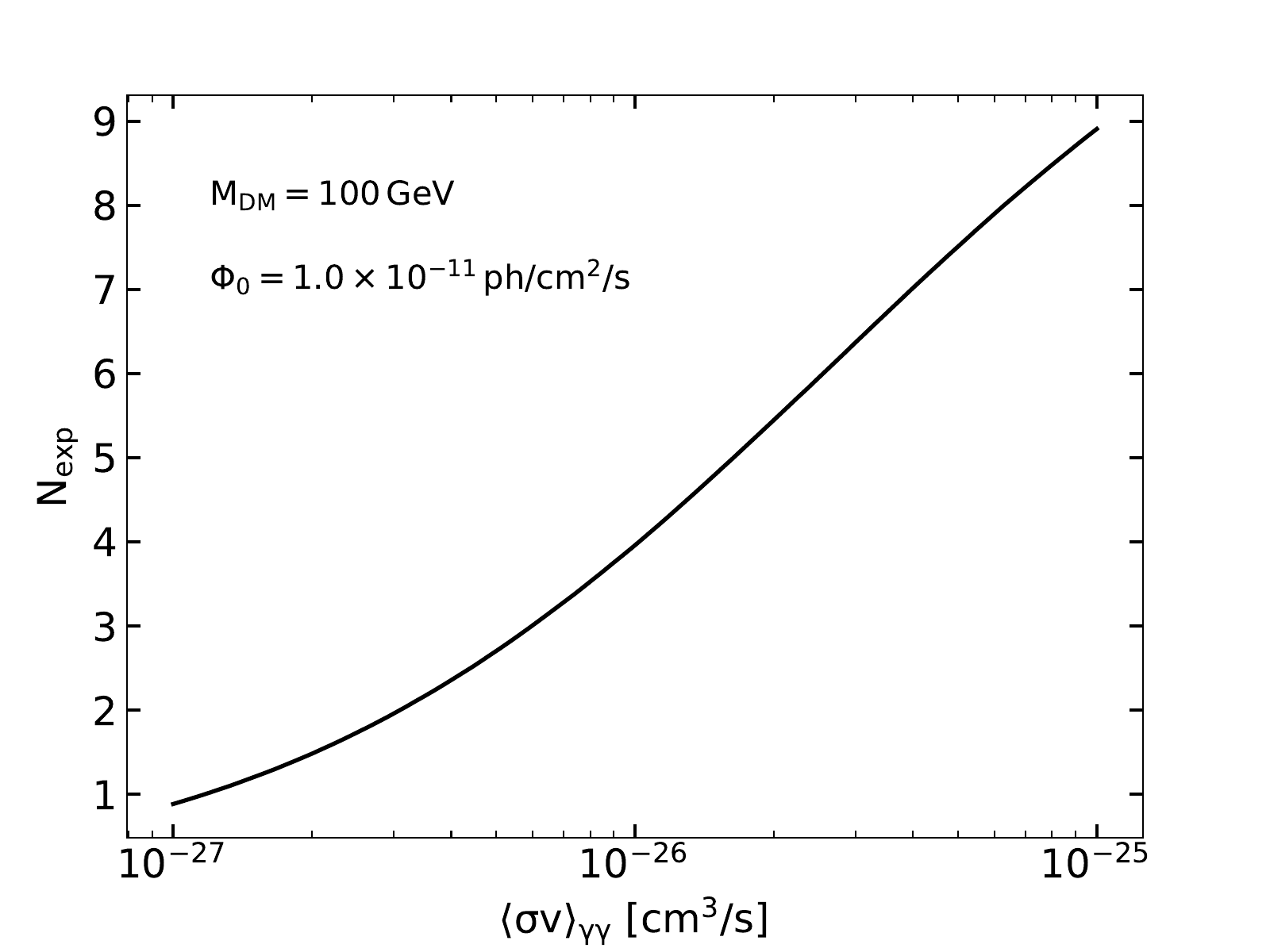}
\caption{{\it Left panel}: The line signal detection threshold $\Phi_0$ as a function of the line energy $E_\gamma$. {\it Right panel}: The expected number of subhalos that can yield line signal significantly detectable by Fermi-LAT as a function of the line signal cross section assuming a DM mass of 100 GeV and a detection threshold of $1.0\times10^{-11}\,{\rm ph/cm^2/s}$. Note that $\Phi_0$, $M_{\rm DM}$ and $\left<\sigma{v}\right>$ are actually three degenerate parameters, the curve in the right panel applies to any values of $\Phi_0$ and $M_{\rm DM}$.}
\label{fig:phi0}
\end{figure*}

No significant line signal ($TS>25$) is found in our analysis and we can set limits on the DM cross section of annihilating to gamma-rays, $\left<\sigma{v}\right>_{\gamma\gamma}$. The basic idea is that, higher cross section may lead to brighter gamma-ray annihilation flux that more subhalos far from us can be detected \cite{fermi12dmsh,berlin14,bertoni15dmsh,schoonenberg16dmsh,hooper16dmsh,calore16dmsh}. The number of expected observable subhalos $N_{\rm exp}$ is therefore proportional to the cross section.
According to poisson statistics, for a model of $N_{\rm exp}$ observable subhalos, the probability distribution function of the number of detected subhalos, $N_{\rm obs}$, is 
\begin{equation}
p(N_{\rm obs}|N_{\rm exp})=\frac{(N_{\rm exp})^{N_{\rm obs}}\exp^{-N_{\rm exp}}}{N_{\rm obs}!}.
\end{equation}
Thus for a given number of real observed subhalos $N'_{\rm obs}$, the 95\% upper limit of $N_{\rm exp}$ corresponds to the one make 
\begin{equation}
\int_{N_{\rm obs}>N'_{\rm obs}}{p(N_{\rm obs}|N_{\rm exp})}{\rm d}N_{\rm obs}>0.95.
\end{equation}
Since we do not find any line-like excesses with $TS>25$, we set $N'_{\rm obs}=0$, leading to $N_{\rm exp}<3$ at 95\% confidence level.

Here we use the expression derived in Ref. \cite{hooper16dmsh} (hereafter H16) to give the predicted numbers of the observable DM subhalos,
\begin{eqnarray}
N_\text{exp} &=& \Omega \int \int \int \int D^2 \, \frac{dN}{dMdV} \,  \frac{dP}{d\gamma} \,\frac{dP}{dR_b} \, \nonumber\\
&&\Theta[\Phi_{\gamma}(M, D, R_b, \gamma)-\Phi_0]\, dM \, dD \,  dR_b \, d\gamma,  \nonumber \\
\label{eq:npred}
\end{eqnarray}
where $D$ and $M$ are the distance and mass of subhalo, respectively. The gamma-ray flux of the line signal generated in a given subhalo is
\begin{equation}
\Phi_{\gamma} = \frac{{\langle \sigma v \rangle}_{\gamma\gamma}}{4 \pi m^2_\chi D^2} \int \rho^2(r) \, dV.
\label{eq:flux}
\end{equation}
For the DM distribution in the subhalo $\rho(r)$, following H16, a density profile of power law with exponetial cutoff (PLE) is adopted rather than the Navarro-Frenk-White \cite{navarro97nfw} one,
\begin{equation}
\rho(r) = \frac{\rho_0}{r^{\gamma}} \, \exp\left(-\frac{r}{R_b}\right)\,.
\label{eq:dmprofile}
\end{equation}
It is found that a PLE density profile can better match the characteristics found in the VL-II and ELVIS simulations considering the effects of tidal stripping \cite{hooper16dmsh}. In Eq. (\ref{eq:npred}), the $dN/dMdV$, $dP/d\gamma$ and $dP/dR_b$ are subhalo distribution and the distributions of the values of $\gamma$ and $R_b$ near the Earth's location. For these distributions we also utilize the formulae reported in H16, which are presented in Appendix \ref{distribution} as well. When deriving the distributions, their dependence on both the subhalo mass and the location relative to the galactic center has been taken into account \cite{hooper16dmsh}. Please note that these distributions in the integrand of Eq. (\ref{eq:npred}) are only valid in the local environment; especially, to simplify the calculation, a uniform subhalo number density ($dN/dV\propto$const) is assumed following H16. We thus consider only the subhalos within the distance of $5\,{\rm kpc}$ \footnote{The bounds of the integral in Eq. (\ref{eq:npred}) for $M$, $R_b$ and $\gamma$ are [$10^5$, $10^{10}$] $M_\odot$, [0, 5] kpc and [0, 1.45], respectively.}. Subhalos at larger distances may also be detectable, our choice of $D_{\rm max}$ will lead to relatively conservative results.

The $\Phi_0$ in Eq. (\ref{eq:npred}) denotes the flux threshold above which the line signals will be significantly detected. Since no line signal is found with $TS>25$, we make use of the Monte Carlo simulation to derive the $\Phi_0$.
We model the Fermi-LAT observation spectrum averaged over all the sky (excluding the Galactic plane and the regions around bright gamma-ray sources, see below) with a PLE function and use this PLE spectrum to approximate the backgrounds in our line searches. Based on this PLE background spectrum, we generate pseudo photons in the 2$^\circ$ ROI. Besides, a line-like component is superposed onto the background, the profile of which is the energy dispersion function of the Fermi-LAT data used in this work. We apply the same searching procedure as that in Sec. \ref{sec2} on these pseudo data, and derive corresponding TS value of the line component. By varying the flux $\Phi$ of the input line component, for each $E_\gamma$ we determine the threshold above which the line component has a TS value greater than 25. We perform 100 Monte Carlo simulations and adopt the median value of the thresholds as the $\Phi_0$. The resultant $\Phi_0$ curve is shown in the left panel of Figure \ref{fig:phi0}.

\begin{figure*}[!t]
\includegraphics[width=0.55\textwidth]{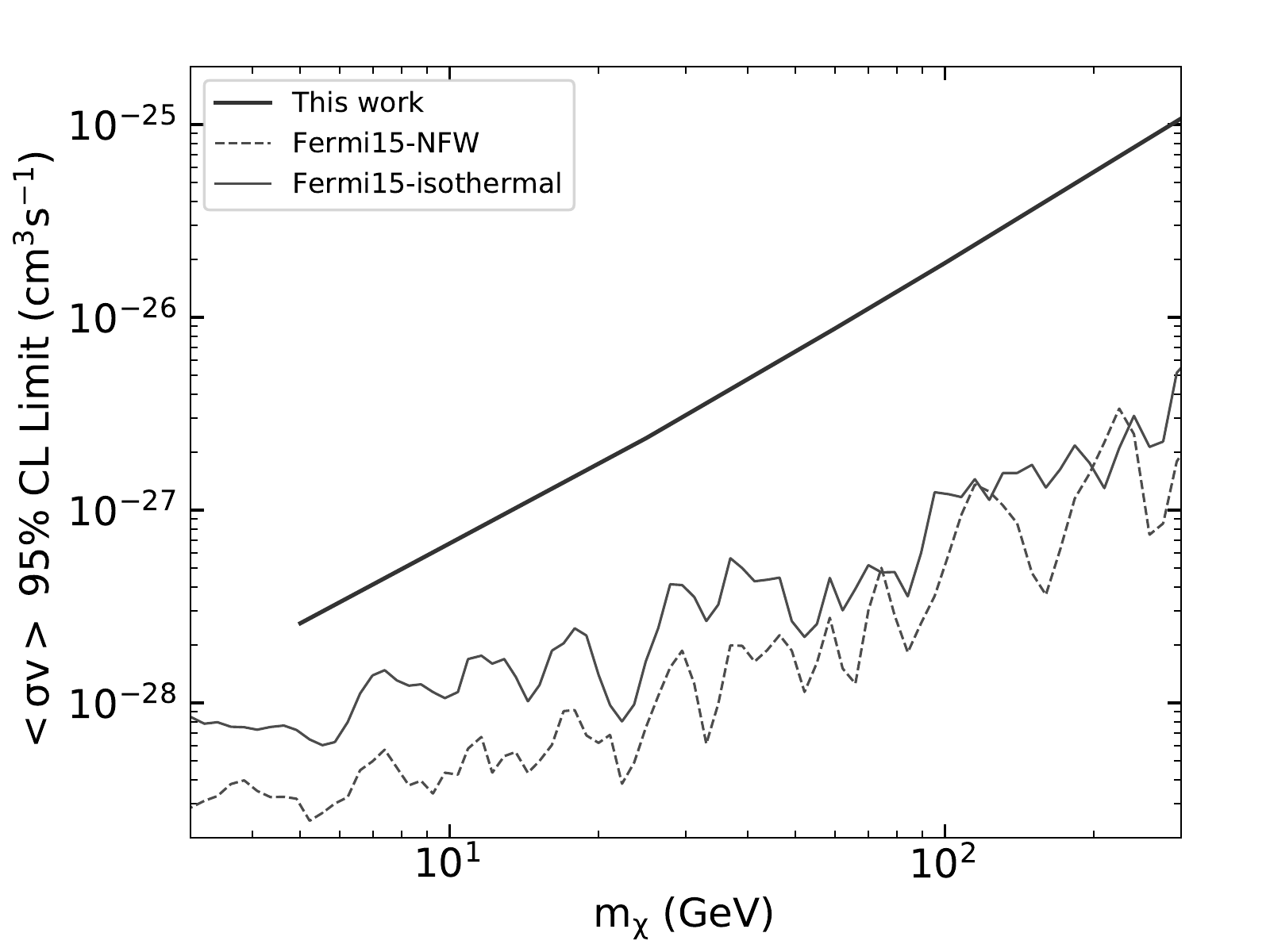}
\caption{The 95\% confidence level upper limits on the cross sections of DM annihilating into double $\gamma$-rays derived in our analysis. As a comparison, we also plot the constraints set by the Fermi-LAT observations of the Galactic central regions \cite{fermi15line}.}
\label{fig:limits}
\end{figure*}

The background gamma-ray emissions in the Galactic plane region and near bright gamma-ray point sources are much stronger than that in other regions, thus lowering the detectability of a line signal from subhalo. In this section, for both calculating the observed solid angle $\Omega$ and deriving the flux threshold $\Phi_0$, the regions of $|b|<20^\circ$ and those within $2^\circ$ around the 100 most bright point sources in 3FGL \cite{fermi15_3fgl} are excluded.

With the elements described above, we can calculate the expected number of subhalos, $N_{\rm exp}$, that can yield line signals significantly detectable by Fermi-LAT. Assuming a detection threshold of $\Phi_0=1.0\times10^{-11}\,{\rm ph\,cm^{-2}\,s^{-1}}$, the $N_{\rm exp}$ as a function of cross section for 100 GeV DM is shown in the right panel of Figure \ref{fig:phi0}. We apply poisson statistics to the $N_{\rm exp}$ (i.e., requiring $N_{\rm exp}<3$) to place a 95\% upper limit on the annihilation cross section for a given value of the DM mass. The obtained constraints are shown in Figure \ref{fig:limits}. As a comparison, also plotted are the constraints derived based on the Fermi-LAT observation towards the regions around the Galactic center \cite{fermi15line}. We find that our constraints here are not competitive with these Galactic ones (thin solid line for the isothermal density profile and dashed line for the NFW).
%However, better than the one in Ref. \cite{Liang17line3} (dotted line), which is also derived based on the subhalo properties.
We would like to emphasize that in the calculation we have considered only the subhalos within 5 kpc. The current constraints would be improved by including the subhalos farther away.

\section{Summary}
In this work, we have analyzed the Fermi-LAT data to blindly search for the potential line signals originated from anywhere of the sky.
We make use of the sliding window technique to perform unbinned likelihood fittings in 49152 ROIs, which cover the whole sky. We did not find line signal with $TS>25$. However, line-like excesses with $TS>16$ appear in the spectra of 50 regions. After the trial factor correction, the highest global significance among these excesses is only $0.54\sigma$. These excesses are most likely originated from statistic fluctuations. In any case, the possibility that very a few of them come from DM annihilation can not be excluded. 
%If the same regions are observed by other/future gamma-ray observatory, 
If one can analyze the data observed by other/future survey mode gamma-ray observatories in the same regions,
their origin (DM or statistical fluctuation) may be identified. We thus suggest that these regions are worth further attention. All these regions have been presented in Appendix \ref{list}.

The DM particles that generate the line signals may simultaneously annihilate through
other channels, thus leading to counterpart gamma-ray emission (either continuum emission
at lower energies or a second gamma-ray line). If detected, these counterparts provide indications of DM origin of the line signals. In Section \ref{sec4}, we have attempted to search for the counterpart gamma-ray emissions for the line-like excesses in Table \ref{tb:50roi} by analyzing the Fermi-LAT
unassociated point sources within selected ROIs (for continuum emission) or by examining the significances of the second lines at specific energies. No evidence of the counterparts is found in the analyses.

Some previous works have pointed out that the number of DM subhalo candidates can be used to place constraints on the DM cross section. In our analysis, we don't find any significant line signal with $TS>25$, then the number of observed subhalo is zero.
Based on this, we derive the expected number of subhalos as a function of the cross section of DM annihilating to gamma-ray lines, and then set constraints on the latter. We found that the constraints obtained here are weaker than those given according to the Fermi-LAT observations towards the Galactic central region. Nevertheless our work offers a novel approach to support these previous constraints independently.

Finally, we would point out that some other on-orbit or proposed space borne gamma-ray telescopes, such as DAMPE \cite{dampe}, Gamma-400 \cite{galper14gamma400} and HERD \cite{zhang14herd}, all of which have significantly better energy resolution comparing to Fermi-LAT, will contribute significantly to the gamma-ray line search and may help examining the origin of the line-like excesses in this work.

\begin{acknowledgments}
We thank Samuel J. Witte for helpful discussion on the calculation of Eq. (\ref{eq:npred}). This work is supported in part by the National Key Research and Development
Program of China (No. 2016YFA0400200), the National Natural Science Foundation
of China (Nos. 11525313, 11722328, 11773075, U1738210, U1738136).
\end{acknowledgments}

\bibliographystyle{apsrev4-1-lyf}
\bibliography{subhalo_line}

%merlin.mbs apsrev4-1.bst 2010-07-25 4.21a (PWD, AO, DPC) hacked
%Control: key (0)
%Control: author (72) initials jnrlst
%Control: editor formatted (1) identically to author
%Control: production of article title (1) required
%Control: page (0) single
%Control: year (1) truncated
%Control: production of eprint (0) enabled
\begin{thebibliography}{41}%
\makeatletter
\providecommand \@ifxundefined [1]{%
 \@ifx{#1\undefined}
}%
\providecommand \@ifnum [1]{%
 \ifnum #1\expandafter \@firstoftwo
 \else \expandafter \@secondoftwo
 \fi
}%
\providecommand \@ifx [1]{%
 \ifx #1\expandafter \@firstoftwo
 \else \expandafter \@secondoftwo
 \fi
}%
\providecommand \natexlab [1]{#1}%
\providecommand \enquote  [1]{``#1''}%
\providecommand \bibnamefont  [1]{#1}%
\providecommand \bibfnamefont [1]{#1}%
\providecommand \citenamefont [1]{#1}%
\providecommand \href@noop [0]{\@secondoftwo}%
\providecommand \href [0]{\begingroup \@sanitize@url \@href}%
\providecommand \@href[1]{\@@startlink{#1}\@@href}%
\providecommand \@@href[1]{\endgroup#1\@@endlink}%
\providecommand \@sanitize@url [0]{\catcode `\\12\catcode `\$12\catcode
  `\&12\catcode `\#12\catcode `\^12\catcode `\_12\catcode `\%12\relax}%
\providecommand \@@startlink[1]{}%
\providecommand \@@endlink[0]{}%
\providecommand \url  [0]{\begingroup\@sanitize@url \@url }%
\providecommand \@url [1]{\endgroup\@href {#1}{\urlprefix }}%
\providecommand \urlprefix  [0]{URL }%
\providecommand \Eprint [0]{\href }%
\providecommand \doibase [0]{http://dx.doi.org/}%
\providecommand \selectlanguage [0]{\@gobble}%
\providecommand \bibinfo  [0]{\@secondoftwo}%
\providecommand \bibfield  [0]{\@secondoftwo}%
\providecommand \translation [1]{[#1]}%
\providecommand \BibitemOpen [0]{}%
\providecommand \bibitemStop [0]{}%
\providecommand \bibitemNoStop [0]{.\EOS\space}%
\providecommand \EOS [0]{\spacefactor3000\relax}%
\providecommand \BibitemShut  [1]{\csname bibitem#1\endcsname}%
\let\auto@bib@innerbib\@empty
%</preamble>
\bibitem [{\citenamefont {{Atwood}}\ \emph {{\it et~al.}}(2009)\citenamefont
  {{Atwood}} {\it et~al.}}]{atwood09LAT}%
  \BibitemOpen
  \bibfield  {author} {\bibinfo {author} {\bibfnamefont {W.~B.}\ \bibnamefont
  {{Atwood}}}  {\it et~al.} (\bibinfo {collaboration} {Fermi LAT}
  Collaboration),\ }\bibfield  {title} {\enquote {\bibinfo {title} {{The Large
  Area Telescope on the Fermi Gamma-Ray Space Telescope Mission}},}\ }\href
  {\doibase 10.1088/0004-637X/697/2/1071} {\bibfield  {journal} {\bibinfo
  {journal} {Astrophys. J.}\ }\textbf {\bibinfo {volume} {697}},\ \bibinfo
  {pages} {1071} (\bibinfo {year} {2009})},\ \Eprint
  {http://arxiv.org/abs/0902.1089}{arXiv:0902.1089}\BibitemShut {NoStop}%
\bibitem [{\citenamefont {{Chang}}\ \emph {{\it et~al.}}(2017)\citenamefont
  {{Chang}}, \citenamefont {{Ambrosi}}, \citenamefont {{An}}, \citenamefont
  {{Asfandiyarov}}, \citenamefont {{Azzarello}}, \citenamefont {{Bernardini}},
  \citenamefont {{Bertucci}}, \citenamefont {{Cai}}, \citenamefont
  {{Caragiulo}}, \citenamefont {{Chen}}, \citenamefont {{Chen}}, \citenamefont
  {{Chen}}, \citenamefont {{Chen}}, \citenamefont {{Cui}}, \citenamefont
  {{Cui}}, \citenamefont {{D'Amone}}, \citenamefont {{De Benedittis}},
  \citenamefont {{De Mitri}}, \citenamefont {{Di Santo}}, \citenamefont
  {{Dong}}, \citenamefont {{Dong}}, \citenamefont {{Dong}}, \citenamefont
  {{Dong}}, \citenamefont {{Donvito}}, \citenamefont {{Droz}}, \citenamefont
  {{Duan}}, \citenamefont {{Duan}}, \citenamefont {{Duranti}}, \citenamefont
  {{D'Urso}}, \citenamefont {{Fan}}, \citenamefont {{Fan}}, \citenamefont
  {{Fang}}, \citenamefont {{Feng}}, \citenamefont {{Feng}}, \citenamefont
  {{Fusco}}, \citenamefont {{Gallo}}, \citenamefont {{Gan}}, \citenamefont
  {{Gan}}, \citenamefont {{Gao}}, \citenamefont {{Gao}}, \citenamefont
  {{Gargano}}, \citenamefont {{Gong}}, \citenamefont {{Gong}}, \citenamefont
  {{Guo}}, \citenamefont {{Hu}}, \citenamefont {{Huang}}, \citenamefont
  {{Huang}}, \citenamefont {{Ionica}}, \citenamefont {{Jiang}}, \citenamefont
  {{Jiang}}, \citenamefont {{Jin}}, \citenamefont {{Kong}}, \citenamefont
  {{Lei}}, \citenamefont {{Li}}, \citenamefont {{Li}}, \citenamefont {{Li}},
  \citenamefont {{Li}}, \citenamefont {{Liang}}, \citenamefont {{Liang}},
  \citenamefont {{Liao}}, \citenamefont {{Liu}}, \citenamefont {{Liu}},
  \citenamefont {{Liu}}, \citenamefont {{Liu}}, \citenamefont {{Liu}},
  \citenamefont {{Liu}}, \citenamefont {{Liu}}, \citenamefont {{Loparco}},
  \citenamefont {{L{\"u}}}, \citenamefont {{Ma}}, \citenamefont {{Ma}},
  \citenamefont {{Ma}}, \citenamefont {{Ma}}, \citenamefont {{Ma}},
  \citenamefont {{Ma}}, \citenamefont {{Marsella}}, \citenamefont
  {{Mazziotta}}, \citenamefont {{Mo}}, \citenamefont {{Miao}}, \citenamefont
  {{Niu}}, \citenamefont {{Pohl}}, \citenamefont {{Peng}}, \citenamefont
  {{Peng}}, \citenamefont {{Qiao}}, \citenamefont {{Rao}}, \citenamefont
  {{Salinas}}, \citenamefont {{Shang}}, \citenamefont {{Shen}}, \citenamefont
  {{Shen}}, \citenamefont {{Shen}}, \citenamefont {{Song}}, \citenamefont
  {{Su}}, \citenamefont {{Su}}, \citenamefont {{Sun}}, \citenamefont {{Surdo}},
  \citenamefont {{Teng}}, \citenamefont {{Tian}}, \citenamefont {{Tykhonov}},
  \citenamefont {{Vagelli}}, \citenamefont {{Vitillo}}, \citenamefont {{Wang}},
  \citenamefont {{Wang}}, \citenamefont {{Wang}}, \citenamefont {{Wang}},
  \citenamefont {{Wang}}, \citenamefont {{Wang}}, \citenamefont {{Wang}},
  \citenamefont {{Wang}}, \citenamefont {{Wang}}, \citenamefont {{Wang}},
  \citenamefont {{Wang}}, \citenamefont {{Wang}}, \citenamefont {{Wang}},
  \citenamefont {{Wen}}, \citenamefont {{Wang}}, \citenamefont {{Wei}},
  \citenamefont {{Wei}}, \citenamefont {{Wei}}, \citenamefont {{Wu}},
  \citenamefont {{Wu}}, \citenamefont {{Wu}}, \citenamefont {{Wu}},
  \citenamefont {{Xi}}, \citenamefont {{Xia}}, \citenamefont {{Xin}},
  \citenamefont {{Xu}}, \citenamefont {{Xu}}, \citenamefont {{Xu}},
  \citenamefont {{Xue}}, \citenamefont {{Yang}}, \citenamefont {{Yang}},
  \citenamefont {{Yang}}, \citenamefont {{Yang}}, \citenamefont {{Yang}},
  \citenamefont {{Yao}}, \citenamefont {{Yu}}, \citenamefont {{Yuan}},
  \citenamefont {{Yue}}, \citenamefont {{Zang}}, \citenamefont {{Zhang}},
  \citenamefont {{Zhang}}, \citenamefont {{Zhang}}, \citenamefont {{Zhang}},
  \citenamefont {{Zhang}}, \citenamefont {{Zhang}}, \citenamefont {{Zhang}},
  \citenamefont {{Zhang}}, \citenamefont {{Zhang}}, \citenamefont {{Zhang}},
  \citenamefont {{Zhang}}, \citenamefont {{Zhang}}, \citenamefont {{Zhang}},
  \citenamefont {{Zhang}}, \citenamefont {{Zhang}}, \citenamefont {{Zhang}},
  \citenamefont {{Zhang}}, \citenamefont {{Zhao}}, \citenamefont {{Zhao}},
  \citenamefont {{Zhao}}, \citenamefont {{Zhou}}, \citenamefont {{Zhou}},
  \citenamefont {{Zhu}}, \citenamefont {{Zhu}},\ and\ \citenamefont
  {{Zimmer}}}]{dampe}%
  \BibitemOpen
  \bibfield  {author} {\bibinfo {author} {\bibfnamefont {J.}~\bibnamefont
  {{Chang}}} {\it et~al.},\ }\bibfield  {title} {\enquote {\bibinfo {title}
  {{The DArk Matter Particle Explorer mission}},}\ }\href {\doibase
  10.1016/j.astropartphys.2017.08.005} {\bibfield  {journal} {\bibinfo
  {journal} {Astroparticle Physics}\ }\textbf {\bibinfo {volume} {95}},\
  \bibinfo {pages} {6} (\bibinfo {year} {2017})},\ \Eprint
  {http://arxiv.org/abs/1706.08453}{arXiv:1706.08453}\BibitemShut {NoStop}%
\bibitem [{\citenamefont {{Pullen}}\ \emph {{\it et~al.}}(2007)\citenamefont
  {{Pullen}}, \citenamefont {{Chary}},\ and\ \citenamefont
  {{Kamionkowski}}}]{pullen07EGRETline}%
  \BibitemOpen
  \bibfield  {author} {\bibinfo {author} {\bibfnamefont {A.~R.}\ \bibnamefont
  {{Pullen}}}, \bibinfo {author} {\bibfnamefont {R.-R.}\ \bibnamefont
  {{Chary}}}, and\ \bibinfo {author} {\bibfnamefont {M.}~\bibnamefont
  {{Kamionkowski}}},\ }\bibfield  {title} {\enquote {\bibinfo {title} {{Search
  with EGRET for a gamma ray line from the Galactic center}},}\ }\href
  {\doibase 10.1103/PhysRevD.76.063006} {\bibfield  {journal} {\bibinfo
  {journal} {\prd}\ }\textbf {\bibinfo {volume} {76}},\ \bibinfo {eid} {063006}
  (\bibinfo {year} {2007})},\ \Eprint
  {http://arxiv.org/abs/astro-ph/0610295}{astro-ph/0610295}\BibitemShut
  {NoStop}%
\bibitem [{\citenamefont {{Abdo}}\ \emph {{\it et~al.}}(2010)\citenamefont
  {{Abdo}}, \citenamefont {{Ackermann}}, \citenamefont {{Ajello}},
  \citenamefont {{Atwood}}, \citenamefont {{Baldini}}, \citenamefont
  {{Ballet}}, \citenamefont {{Barbiellini}}, \citenamefont {{Bastieri}},
  \citenamefont {{Bechtol}}, \citenamefont {{Bellazzini}}, \citenamefont
  {{Berenji}}, \citenamefont {{Bloom}}, \citenamefont {{Bonamente}},
  \citenamefont {{Borgland}}, \citenamefont {{Bouvier}}, \citenamefont
  {{Bregeon}}, \citenamefont {{Brez}}, \citenamefont {{Brigida}}, \citenamefont
  {{Bruel}}, \citenamefont {{Burnett}}, \citenamefont {{Buson}}, \citenamefont
  {{Caliandro}}, \citenamefont {{Cameron}}, \citenamefont {{Caraveo}},
  \citenamefont {{Carrigan}}, \citenamefont {{Casandjian}}, \citenamefont
  {{Cecchi}}, \citenamefont {{{\c C}elik}}, \citenamefont {{Chekhtman}},
  \citenamefont {{Chiang}}, \citenamefont {{Ciprini}}, \citenamefont {{Claus}},
  \citenamefont {{Cohen-Tanugi}}, \citenamefont {{Conrad}}, \citenamefont
  {{Dermer}}, \citenamefont {{de Angelis}}, \citenamefont {{de Palma}},
  \citenamefont {{Digel}}, \citenamefont {{Do Couto E Silva}}, \citenamefont
  {{Drell}}, \citenamefont {{Drlica-Wagner}}, \citenamefont {{Dubois}},
  \citenamefont {{Dumora}}, \citenamefont {{Edmonds}}, \citenamefont {{Essig}},
  \citenamefont {{Farnier}}, \citenamefont {{Favuzzi}}, \citenamefont
  {{Fegan}}, \citenamefont {{Focke}}, \citenamefont {{Fortin}}, \citenamefont
  {{Frailis}}, \citenamefont {{Fukazawa}}, \citenamefont {{Funk}},
  \citenamefont {{Fusco}}, \citenamefont {{Gargano}}, \citenamefont
  {{Gasparrini}}, \citenamefont {{Gehrels}}, \citenamefont {{Germani}},
  \citenamefont {{Giglietto}}, \citenamefont {{Giordano}}, \citenamefont
  {{Glanzman}}, \citenamefont {{Godfrey}}, \citenamefont {{Grenier}},
  \citenamefont {{Grove}}, \citenamefont {{Guillemot}}, \citenamefont
  {{Guiriec}}, \citenamefont {{Gustafsson}}, \citenamefont {{Hadasch}},
  \citenamefont {{Harding}}, \citenamefont {{Horan}}, \citenamefont {{Hughes}},
  \citenamefont {{Jackson}}, \citenamefont {{J{\'o}hannesson}}, \citenamefont
  {{Johnson}}, \citenamefont {{Johnson}}, \citenamefont {{Johnson}},
  \citenamefont {{Kamae}}, \citenamefont {{Katagiri}}, \citenamefont
  {{Kataoka}}, \citenamefont {{Kawai}}, \citenamefont {{Kerr}}, \citenamefont
  {{Kn{\"o}dlseder}}, \citenamefont {{Kuss}}, \citenamefont {{Lande}},
  \citenamefont {{Latronico}}, \citenamefont {{Llena Garde}}, \citenamefont
  {{Longo}}, \citenamefont {{Loparco}}, \citenamefont {{Lott}}, \citenamefont
  {{Lovellette}}, \citenamefont {{Lubrano}}, \citenamefont {{Makeev}},
  \citenamefont {{Mazziotta}}, \citenamefont {{McEnery}}, \citenamefont
  {{Meurer}}, \citenamefont {{Michelson}}, \citenamefont {{Mitthumsiri}},
  \citenamefont {{Mizuno}}, \citenamefont {{Moiseev}}, \citenamefont {{Monte}},
  \citenamefont {{Monzani}}, \citenamefont {{Morselli}}, \citenamefont
  {{Moskalenko}}, \citenamefont {{Murgia}}, \citenamefont {{Nolan}},
  \citenamefont {{Norris}}, \citenamefont {{Nuss}}, \citenamefont {{Ohsugi}},
  \citenamefont {{Omodei}}, \citenamefont {{Orlando}}, \citenamefont {{Ormes}},
  \citenamefont {{Ozaki}}, \citenamefont {{Paneque}}, \citenamefont
  {{Panetta}}, \citenamefont {{Parent}}, \citenamefont {{Pelassa}},
  \citenamefont {{Pepe}}, \citenamefont {{Pesce-Rollins}}, \citenamefont
  {{Piron}}, \citenamefont {{Rain{\`o}}}, \citenamefont {{Rando}},
  \citenamefont {{Razzano}}, \citenamefont {{Reimer}}, \citenamefont
  {{Reimer}}, \citenamefont {{Reposeur}}, \citenamefont {{Ripken}},
  \citenamefont {{Ritz}}, \citenamefont {{Rodriguez}}, \citenamefont {{Roth}},
  \citenamefont {{Sadrozinski}}, \citenamefont {{Sander}}, \citenamefont
  {{Parkinson}}, \citenamefont {{Scargle}}, \citenamefont {{Schalk}},
  \citenamefont {{Sellerholm}}, \citenamefont {{Sgr{\`o}}}, \citenamefont
  {{Siskind}}, \citenamefont {{Smith}}, \citenamefont {{Smith}}, \citenamefont
  {{Spandre}}, \citenamefont {{Spinelli}}, \citenamefont {{Starck}},
  \citenamefont {{Strickman}}, \citenamefont {{Suson}}, \citenamefont
  {{Tajima}}, \citenamefont {{Takahashi}}, \citenamefont {{Tanaka}},
  \citenamefont {{Thayer}}, \citenamefont {{Thayer}}, \citenamefont
  {{Tibaldo}}, \citenamefont {{Torres}}, \citenamefont {{Uchiyama}},
  \citenamefont {{Usher}}, \citenamefont {{Vasileiou}}, \citenamefont
  {{Vilchez}}, \citenamefont {{Vitale}}, \citenamefont {{Waite}}, \citenamefont
  {{Wang}}, \citenamefont {{Winer}}, \citenamefont {{Wood}}, \citenamefont
  {{Ylinen}}, \citenamefont {{Ziegler}},\ and\ \citenamefont {{Fermi LAT
  Collaboration}}}]{fermi10line1}%
  \BibitemOpen
  \bibfield  {author} {\bibinfo {author} {\bibfnamefont {A.~A.}\ \bibnamefont
  {{Abdo}}} {\it et~al.},\ }\bibfield  {title} {\enquote {\bibinfo {title}
  {{Fermi Large Area Telescope Search for Photon Lines from 30 to 200 GeV and
  Dark Matter Implications}},}\ }\href {\doibase
  10.1103/PhysRevLett.104.091302} {\bibfield  {journal} {\bibinfo  {journal}
  {Physical Review Letters}\ }\textbf {\bibinfo {volume} {104}},\ \bibinfo
  {eid} {091302} (\bibinfo {year} {2010})},\ \Eprint
  {http://arxiv.org/abs/1001.4836}{arXiv:1001.4836}\BibitemShut {NoStop}%
\bibitem [{\citenamefont {{Ackermann}}\ \emph {{\it
  et~al.}}(2012{\natexlab{a}})\citenamefont {{Ackermann}}, \citenamefont
  {{Ajello}}, \citenamefont {{Albert}}, \citenamefont {{Baldini}},
  \citenamefont {{Barbiellini}}, \citenamefont {{Bechtol}}, \citenamefont
  {{Bellazzini}}, \citenamefont {{Berenji}}, \citenamefont {{Blandford}},
  \citenamefont {{Bloom}}, \citenamefont {{Bonamente}}, \citenamefont
  {{Borgland}}, \citenamefont {{Brigida}}, \citenamefont {{Buehler}},
  \citenamefont {{Buson}}, \citenamefont {{Caliandro}}, \citenamefont
  {{Cameron}}, \citenamefont {{Caraveo}}, \citenamefont {{Casandjian}},
  \citenamefont {{Cecchi}}, \citenamefont {{Charles}}, \citenamefont
  {{Chekhtman}}, \citenamefont {{Chiang}}, \citenamefont {{Ciprini}},
  \citenamefont {{Claus}}, \citenamefont {{Cohen-Tanugi}}, \citenamefont
  {{Conrad}}, \citenamefont {{D'Ammando}}, \citenamefont {{de Palma}},
  \citenamefont {{Dermer}}, \citenamefont {{do Couto e Silva}}, \citenamefont
  {{Drell}}, \citenamefont {{Drlica-Wagner}}, \citenamefont {{Edmonds}},
  \citenamefont {{Essig}}, \citenamefont {{Favuzzi}}, \citenamefont {{Fegan}},
  \citenamefont {{Focke}}, \citenamefont {{Fukazawa}}, \citenamefont {{Funk}},
  \citenamefont {{Fusco}}, \citenamefont {{Gargano}}, \citenamefont
  {{Gasparrini}}, \citenamefont {{Germani}}, \citenamefont {{Giglietto}},
  \citenamefont {{Giordano}}, \citenamefont {{Giroletti}}, \citenamefont
  {{Glanzman}}, \citenamefont {{Godfrey}}, \citenamefont {{Grenier}},
  \citenamefont {{Guiriec}}, \citenamefont {{Gustafsson}}, \citenamefont
  {{Hadasch}}, \citenamefont {{Hayashida}}, \citenamefont {{Horan}},
  \citenamefont {{Hughes}}, \citenamefont {{Kamae}}, \citenamefont
  {{Kn{\"o}dlseder}}, \citenamefont {{Kuss}}, \citenamefont {{Lande}},
  \citenamefont {{Lionetto}}, \citenamefont {{Llena Garde}}, \citenamefont
  {{Longo}}, \citenamefont {{Loparco}}, \citenamefont {{Lovellette}},
  \citenamefont {{Lubrano}}, \citenamefont {{Mazziotta}}, \citenamefont
  {{Michelson}}, \citenamefont {{Mitthumsiri}}, \citenamefont {{Mizuno}},
  \citenamefont {{Moiseev}}, \citenamefont {{Monte}}, \citenamefont
  {{Monzani}}, \citenamefont {{Morselli}}, \citenamefont {{Moskalenko}},
  \citenamefont {{Murgia}}, \citenamefont {{Naumann-Godo}}, \citenamefont
  {{Norris}}, \citenamefont {{Nuss}}, \citenamefont {{Ohsugi}}, \citenamefont
  {{Okumura}}, \citenamefont {{Orlando}}, \citenamefont {{Ormes}},
  \citenamefont {{Paneque}}, \citenamefont {{Panetta}}, \citenamefont
  {{Pesce-Rollins}}, \citenamefont {{Piron}}, \citenamefont {{Pivato}},
  \citenamefont {{Porter}}, \citenamefont {{Prokhorov}}, \citenamefont
  {{Rain{\`o}}}, \citenamefont {{Rando}}, \citenamefont {{Razzano}},
  \citenamefont {{Reimer}}, \citenamefont {{Roth}}, \citenamefont {{Sbarra}},
  \citenamefont {{Scargle}}, \citenamefont {{Sgr{\`o}}}, \citenamefont
  {{Siskind}}, \citenamefont {{Snyder}}, \citenamefont {{Spinelli}},
  \citenamefont {{Suson}}, \citenamefont {{Takahashi}}, \citenamefont
  {{Tanaka}}, \citenamefont {{Thayer}}, \citenamefont {{Thayer}}, \citenamefont
  {{Tibaldo}}, \citenamefont {{Tinivella}}, \citenamefont {{Torres}},
  \citenamefont {{Tosti}}, \citenamefont {{Troja}}, \citenamefont
  {{Vandenbroucke}}, \citenamefont {{Vasileiou}}, \citenamefont {{Vianello}},
  \citenamefont {{Vitale}}, \citenamefont {{Waite}}, \citenamefont {{Winer}},
  \citenamefont {{Wood}}, \citenamefont {{Yang}},\ and\ \citenamefont
  {{Zimmer}}}]{fermi12line2}%
  \BibitemOpen
  \bibfield  {author} {\bibinfo {author} {\bibfnamefont {M.}~\bibnamefont
  {{Ackermann}}} {\it et~al.},\ }\bibfield  {title} {\enquote {\bibinfo {title}
  {{Fermi LAT search for dark matter in gamma-ray lines and the inclusive
  photon spectrum}},}\ }\href {\doibase 10.1103/PhysRevD.86.022002} {\bibfield
  {journal} {\bibinfo  {journal} {\prd}\ }\textbf {\bibinfo {volume} {86}},\
  \bibinfo {eid} {022002} (\bibinfo {year} {2012}{\natexlab{a}})},\ \Eprint
  {http://arxiv.org/abs/1205.2739}{arXiv:1205.2739}\BibitemShut {NoStop}%
\bibitem [{\citenamefont {{Bringmann}}\ \emph {{\it et~al.}}(2012)\citenamefont
  {{Bringmann}}, \citenamefont {{Huang}}, \citenamefont {{Ibarra}},
  \citenamefont {{Vogl}},\ and\ \citenamefont
  {{Weniger}}}]{bringmann12_130gev}%
  \BibitemOpen
  \bibfield  {author} {\bibinfo {author} {\bibfnamefont {T.}~\bibnamefont
  {{Bringmann}}}, \bibinfo {author} {\bibfnamefont {X.}~\bibnamefont
  {{Huang}}}, \bibinfo {author} {\bibfnamefont {A.}~\bibnamefont {{Ibarra}}},
  \bibinfo {author} {\bibfnamefont {S.}~\bibnamefont {{Vogl}}}, and\ \bibinfo
  {author} {\bibfnamefont {C.}~\bibnamefont {{Weniger}}},\ }\bibfield  {title}
  {\enquote {\bibinfo {title} {{Fermi LAT search for internal bremsstrahlung
  signatures from dark matter annihilation}},}\ }\href {\doibase
  10.1088/1475-7516/2012/07/054} {\bibfield  {journal} {\bibinfo  {journal}
  {\jcap}\ }\textbf {\bibinfo {volume} {7}},\ \bibinfo {eid} {054} (\bibinfo
  {year} {2012})},\ \Eprint
  {http://arxiv.org/abs/1203.1312}{arXiv:1203.1312}\BibitemShut {NoStop}%
\bibitem [{\citenamefont {{Weniger}}(2012)}]{weniger12_130gev}%
  \BibitemOpen
  \bibfield  {author} {\bibinfo {author} {\bibfnamefont {C.}~\bibnamefont
  {{Weniger}}},\ }\bibfield  {title} {\enquote {\bibinfo {title} {{A tentative
  gamma-ray line from Dark Matter annihilation at the Fermi Large Area
  Telescope}},}\ }\href {\doibase 10.1088/1475-7516/2012/08/007} {\bibfield
  {journal} {\bibinfo  {journal} {\jcap}\ }\textbf {\bibinfo {volume} {8}},\
  \bibinfo {eid} {007} (\bibinfo {year} {2012})},\ \Eprint
  {http://arxiv.org/abs/1204.2797}{arXiv:1204.2797}\BibitemShut {NoStop}%
\bibitem [{\citenamefont {{Geringer-Sameth}}\ and\ \citenamefont
  {{Koushiappas}}(2012)}]{geringer12dsphLine}%
  \BibitemOpen
  \bibfield  {author} {\bibinfo {author} {\bibfnamefont {A.}~\bibnamefont
  {{Geringer-Sameth}}} and\ \bibinfo {author} {\bibfnamefont {S.~M.}\
  \bibnamefont {{Koushiappas}}},\ }\bibfield  {title} {\enquote {\bibinfo
  {title} {{Dark matter line search using a joint analysis of dwarf galaxies
  with the Fermi Gamma-ray Space Telescope}},}\ }\href {\doibase
  10.1103/PhysRevD.86.021302} {\bibfield  {journal} {\bibinfo  {journal}
  {\prd}\ }\textbf {\bibinfo {volume} {86}},\ \bibinfo {eid} {021302} (\bibinfo
  {year} {2012})},\ \Eprint
  {http://arxiv.org/abs/1206.0796}{arXiv:1206.0796}\BibitemShut {NoStop}%
\bibitem [{\citenamefont {{Tempel}}\ \emph {{\it et~al.}}(2012)\citenamefont
  {{Tempel}}, \citenamefont {{Hektor}},\ and\ \citenamefont
  {{Raidal}}}]{tempel12_130gev}%
  \BibitemOpen
  \bibfield  {author} {\bibinfo {author} {\bibfnamefont {E.}~\bibnamefont
  {{Tempel}}}, \bibinfo {author} {\bibfnamefont {A.}~\bibnamefont {{Hektor}}},
  and\ \bibinfo {author} {\bibfnamefont {M.}~\bibnamefont {{Raidal}}},\
  }\bibfield  {title} {\enquote {\bibinfo {title} {{Fermi 130 GeV gamma-ray
  excess and dark matter annihilation in sub-haloes and in the Galactic
  centre}},}\ }\href {\doibase 10.1088/1475-7516/2012/09/032} {\bibfield
  {journal} {\bibinfo  {journal} {\jcap}\ }\textbf {\bibinfo {volume} {9}},\
  \bibinfo {eid} {032} (\bibinfo {year} {2012})},\ \Eprint
  {http://arxiv.org/abs/1205.1045}{arXiv:1205.1045}\BibitemShut {NoStop}%
\bibitem [{\citenamefont {{Huang}}\ \emph {{\it et~al.}}(2012)\citenamefont
  {{Huang}}, \citenamefont {{Yuan}}, \citenamefont {{Yin}}, \citenamefont
  {{Bi}},\ and\ \citenamefont {{Chen}}}]{huang12_130gev}%
  \BibitemOpen
  \bibfield  {author} {\bibinfo {author} {\bibfnamefont {X.}~\bibnamefont
  {{Huang}}}, \bibinfo {author} {\bibfnamefont {Q.}~\bibnamefont {{Yuan}}},
  \bibinfo {author} {\bibfnamefont {P.-F.}\ \bibnamefont {{Yin}}}, \bibinfo
  {author} {\bibfnamefont {X.-J.}\ \bibnamefont {{Bi}}}, and\ \bibinfo {author}
  {\bibfnamefont {X.}~\bibnamefont {{Chen}}},\ }\bibfield  {title} {\enquote
  {\bibinfo {title} {{Constraints on the dark matter annihilation scenario of
  Fermi 130 GeV gamma-ray line emission by continuous gamma-rays, Milky Way
  halo, galaxy clusters and dwarf galaxies observations}},}\ }\href {\doibase
  10.1088/1475-7516/2012/11/048} {\bibfield  {journal} {\bibinfo  {journal}
  {\jcap}\ }\textbf {\bibinfo {volume} {11}},\ \bibinfo {eid} {048} (\bibinfo
  {year} {2012})},\ \Eprint
  {http://arxiv.org/abs/1208.0267}{arXiv:1208.0267}\BibitemShut {NoStop}%
\bibitem [{\citenamefont {{Su}}\ and\ \citenamefont
  {{Finkbeiner}}(2012)}]{su2012line}%
  \BibitemOpen
  \bibfield  {author} {\bibinfo {author} {\bibfnamefont {M.}~\bibnamefont
  {{Su}}} and\ \bibinfo {author} {\bibfnamefont {D.~P.}\ \bibnamefont
  {{Finkbeiner}}},\ }\bibfield  {title} {\enquote {\bibinfo {title} {{Strong
  Evidence for Gamma-ray Line Emission from the Inner Galaxy}},}\ }\href@noop
  {} {\bibfield  {journal} {\bibinfo  {journal} {ArXiv e-prints}\ } (\bibinfo
  {year} {2012})},\ \Eprint
  {http://arxiv.org/abs/1206.1616}{arXiv:1206.1616}\BibitemShut {NoStop}%
\bibitem [{\citenamefont {{Ackermann}}\ \emph {{\it et~al.}}(2013)\citenamefont
  {{Ackermann}} {\it et~al.}}]{fermi13line}%
  \BibitemOpen
  \bibfield  {author} {\bibinfo {author} {\bibfnamefont {M.}~\bibnamefont
  {{Ackermann}}}  {\it et~al.} (\bibinfo {collaboration} {Fermi LAT}
  Collaboration),\ }\bibfield  {title} {\enquote {\bibinfo {title} {{Search for
  gamma-ray spectral lines with the Fermi Large Area Telescope and dark matter
  implications}},}\ }\href {\doibase 10.1103/PhysRevD.88.082002} {\bibfield
  {journal} {\bibinfo  {journal} {\prd}\ }\textbf {\bibinfo {volume} {88}},\
  \bibinfo {eid} {082002} (\bibinfo {year} {2013})}\BibitemShut {NoStop}%
\bibitem [{\citenamefont {{Hektor}}\ \emph {{\it et~al.}}(2013)\citenamefont
  {{Hektor}}, \citenamefont {{Raidal}},\ and\ \citenamefont
  {{Tempel}}}]{Hektor2013gcline}%
  \BibitemOpen
  \bibfield  {author} {\bibinfo {author} {\bibfnamefont {A.}~\bibnamefont
  {{Hektor}}}, \bibinfo {author} {\bibfnamefont {M.}~\bibnamefont {{Raidal}}},
  and\ \bibinfo {author} {\bibfnamefont {E.}~\bibnamefont {{Tempel}}},\
  }\bibfield  {title} {\enquote {\bibinfo {title} {{Evidence for Indirect
  Detection of Dark Matter from Galaxy Clusters in Fermi {$\gamma$}-Ray
  Data}},}\ }\href {\doibase 10.1088/2041-8205/762/2/L22} {\bibfield  {journal}
  {\bibinfo  {journal} {\apjl}\ }\textbf {\bibinfo {volume} {762}},\ \bibinfo
  {eid} {L22} (\bibinfo {year} {2013})},\ \Eprint
  {http://arxiv.org/abs/1207.4466}{arXiv:1207.4466}\BibitemShut {NoStop}%
\bibitem [{\citenamefont {{Albert}}\ \emph {{\it et~al.}}(2014)\citenamefont
  {{Albert}}, \citenamefont {{G{\'o}mez-Vargas}}, \citenamefont {{Grefe}},
  \citenamefont {{Mu{\~n}oz}}, \citenamefont {{Weniger}}, \citenamefont
  {{Bloom}}, \citenamefont {{Charles}}, \citenamefont {{Mazziotta}},\ and\
  \citenamefont {{Morselli}}}]{albert14line}%
  \BibitemOpen
  \bibfield  {author} {\bibinfo {author} {\bibfnamefont {A.}~\bibnamefont
  {{Albert}}}, \bibinfo {author} {\bibfnamefont {G.~A.}\ \bibnamefont
  {{G{\'o}mez-Vargas}}}, \bibinfo {author} {\bibfnamefont {M.}~\bibnamefont
  {{Grefe}}}, \bibinfo {author} {\bibfnamefont {C.}~\bibnamefont
  {{Mu{\~n}oz}}}, \bibinfo {author} {\bibfnamefont {C.}~\bibnamefont
  {{Weniger}}}, \bibinfo {author} {\bibfnamefont {E.~D.}\ \bibnamefont
  {{Bloom}}}, \bibinfo {author} {\bibfnamefont {E.}~\bibnamefont {{Charles}}},
  \bibinfo {author} {\bibfnamefont {M.~N.}\ \bibnamefont {{Mazziotta}}}, and\
  \bibinfo {author} {\bibfnamefont {A.}~\bibnamefont {{Morselli}}},\ }\bibfield
   {title} {\enquote {\bibinfo {title} {{Search for 100 MeV to 10 GeV
  {$\gamma$}-ray lines in the Fermi-LAT data and implications for gravitino
  dark matter in the {$\mu$}{$\nu$}SSM}},}\ }\href {\doibase
  10.1088/1475-7516/2014/10/023} {\bibfield  {journal} {\bibinfo  {journal}
  {\jcap}\ }\textbf {\bibinfo {volume} {10}},\ \bibinfo {eid} {023} (\bibinfo
  {year} {2014})},\ \Eprint
  {http://arxiv.org/abs/1406.3430}{arXiv:1406.3430}\BibitemShut {NoStop}%
\bibitem [{\citenamefont {{Ackermann}}\ \emph {{\it et~al.}}(2015)\citenamefont
  {{Ackermann}} {\it et~al.}}]{fermi15line}%
  \BibitemOpen
  \bibfield  {author} {\bibinfo {author} {\bibfnamefont {M.}~\bibnamefont
  {{Ackermann}}}  {\it et~al.} (\bibinfo {collaboration} {Fermi LAT}
  Collaboration),\ }\bibfield  {title} {\enquote {\bibinfo {title} {{Updated
  search for spectral lines from Galactic dark matter interactions with pass 8
  data from the Fermi Large Area Telescope}},}\ }\href {\doibase
  10.1103/PhysRevD.91.122002} {\bibfield  {journal} {\bibinfo  {journal}
  {\prd}\ }\textbf {\bibinfo {volume} {91}},\ \bibinfo {eid} {122002} (\bibinfo
  {year} {2015})}\BibitemShut {NoStop}%
\bibitem [{\citenamefont {{Anderson}}\ \emph {{\it et~al.}}(2016)\citenamefont
  {{Anderson}}, \citenamefont {{Zimmer}}, \citenamefont {{Conrad}},
  \citenamefont {{Gustafsson}}, \citenamefont {{S{\'a}nchez-Conde}},\ and\
  \citenamefont {{Caputo}}}]{anderson16gclsLine}%
  \BibitemOpen
  \bibfield  {author} {\bibinfo {author} {\bibfnamefont {B.}~\bibnamefont
  {{Anderson}}}, \bibinfo {author} {\bibfnamefont {S.}~\bibnamefont
  {{Zimmer}}}, \bibinfo {author} {\bibfnamefont {J.}~\bibnamefont {{Conrad}}},
  \bibinfo {author} {\bibfnamefont {M.}~\bibnamefont {{Gustafsson}}}, \bibinfo
  {author} {\bibfnamefont {M.}~\bibnamefont {{S{\'a}nchez-Conde}}}, and\
  \bibinfo {author} {\bibfnamefont {R.}~\bibnamefont {{Caputo}}},\ }\bibfield
  {title} {\enquote {\bibinfo {title} {{Search for gamma-ray lines towards
  galaxy clusters with the Fermi-LAT}},}\ }\href {\doibase
  10.1088/1475-7516/2016/02/026} {\bibfield  {journal} {\bibinfo  {journal}
  {\jcap}\ }\textbf {\bibinfo {volume} {2}},\ \bibinfo {eid} {026} (\bibinfo
  {year} {2016})},\ \Eprint
  {http://arxiv.org/abs/1511.00014}{arXiv:1511.00014}\BibitemShut {NoStop}%
\bibitem [{\citenamefont {{Liang}}\ \emph {{\it
  et~al.}}(2016{\natexlab{a}})\citenamefont {{Liang}}, \citenamefont {{Shen}},
  \citenamefont {{Li}}, \citenamefont {{Fan}}, \citenamefont {{Huang}},
  \citenamefont {{Lei}}, \citenamefont {{Feng}}, \citenamefont {{Liang}},\ and\
  \citenamefont {{Chang}}}]{liang16gclsLine}%
  \BibitemOpen
  \bibfield  {author} {\bibinfo {author} {\bibfnamefont {Y.-F.}\ \bibnamefont
  {{Liang}}}, \bibinfo {author} {\bibfnamefont {Z.-Q.}\ \bibnamefont {{Shen}}},
  \bibinfo {author} {\bibfnamefont {X.}~\bibnamefont {{Li}}}, \bibinfo {author}
  {\bibfnamefont {Y.-Z.}\ \bibnamefont {{Fan}}}, \bibinfo {author}
  {\bibfnamefont {X.}~\bibnamefont {{Huang}}}, \bibinfo {author} {\bibfnamefont
  {S.-J.}\ \bibnamefont {{Lei}}}, \bibinfo {author} {\bibfnamefont
  {L.}~\bibnamefont {{Feng}}}, \bibinfo {author} {\bibfnamefont {E.-W.}\
  \bibnamefont {{Liang}}}, and\ \bibinfo {author} {\bibfnamefont
  {J.}~\bibnamefont {{Chang}}},\ }\bibfield  {title} {\enquote {\bibinfo
  {title} {{Search for a gamma-ray line feature from a group of nearby galaxy
  clusters with Fermi LAT Pass 8 data}},}\ }\href {\doibase
  10.1103/PhysRevD.93.103525} {\bibfield  {journal} {\bibinfo  {journal}
  {\prd}\ }\textbf {\bibinfo {volume} {93}},\ \bibinfo {eid} {103525} (\bibinfo
  {year} {2016}{\natexlab{a}})},\ \Eprint
  {http://arxiv.org/abs/1602.06527}{arXiv:1602.06527}\BibitemShut {NoStop}%
\bibitem [{\citenamefont {{Profumo}}\ \emph {{\it et~al.}}(2016)\citenamefont
  {{Profumo}}, \citenamefont {{Queiroz}},\ and\ \citenamefont
  {{Yaguna}}}]{Profumo2016line}%
  \BibitemOpen
  \bibfield  {author} {\bibinfo {author} {\bibfnamefont {S.}~\bibnamefont
  {{Profumo}}}, \bibinfo {author} {\bibfnamefont {F.~S.}\ \bibnamefont
  {{Queiroz}}}, and\ \bibinfo {author} {\bibfnamefont {C.~E.}\ \bibnamefont
  {{Yaguna}}},\ }\bibfield  {title} {\enquote {\bibinfo {title} {{Extending
  Fermi-LAT and H.E.S.S. limits on gamma-ray lines from dark matter
  annihilation}},}\ }\href {\doibase 10.1093/mnras/stw1600} {\bibfield
  {journal} {\bibinfo  {journal} {\mnras}\ }\textbf {\bibinfo {volume} {461}},\
  \bibinfo {pages} {3976} (\bibinfo {year} {2016})},\ \Eprint
  {http://arxiv.org/abs/1602.08501}{arXiv:1602.08501}\BibitemShut {NoStop}%
\bibitem [{\citenamefont {{Liang}}\ \emph {{\it
  et~al.}}(2016{\natexlab{b}})\citenamefont {{Liang}}, \citenamefont {{Xia}},
  \citenamefont {{Shen}}, \citenamefont {{Li}}, \citenamefont {{Jiang}},
  \citenamefont {{Yuan}}, \citenamefont {{Fan}}, \citenamefont {{Feng}},
  \citenamefont {{Liang}},\ and\ \citenamefont {{Chang}}}]{liang16dsphLine}%
  \BibitemOpen
  \bibfield  {author} {\bibinfo {author} {\bibfnamefont {Y.-F.}\ \bibnamefont
  {{Liang}}}, \bibinfo {author} {\bibfnamefont {Z.-Q.}\ \bibnamefont {{Xia}}},
  \bibinfo {author} {\bibfnamefont {Z.-Q.}\ \bibnamefont {{Shen}}}, \bibinfo
  {author} {\bibfnamefont {X.}~\bibnamefont {{Li}}}, \bibinfo {author}
  {\bibfnamefont {W.}~\bibnamefont {{Jiang}}}, \bibinfo {author} {\bibfnamefont
  {Q.}~\bibnamefont {{Yuan}}}, \bibinfo {author} {\bibfnamefont {Y.-Z.}\
  \bibnamefont {{Fan}}}, \bibinfo {author} {\bibfnamefont {L.}~\bibnamefont
  {{Feng}}}, \bibinfo {author} {\bibfnamefont {E.-W.}\ \bibnamefont {{Liang}}},
  and\ \bibinfo {author} {\bibfnamefont {J.}~\bibnamefont {{Chang}}},\
  }\bibfield  {title} {\enquote {\bibinfo {title} {{Search for gamma-ray line
  features from Milky Way satellites with Fermi LAT Pass 8 data}},}\ }\href
  {\doibase 10.1103/PhysRevD.94.103502} {\bibfield  {journal} {\bibinfo
  {journal} {\prd}\ }\textbf {\bibinfo {volume} {94}},\ \bibinfo {eid} {103502}
  (\bibinfo {year} {2016}{\natexlab{b}})},\ \Eprint
  {http://arxiv.org/abs/1608.07184}{arXiv:1608.07184}\BibitemShut {NoStop}%
\bibitem [{\citenamefont {{Liang}}\ \emph {{\it et~al.}}(2017)\citenamefont
  {{Liang}}, \citenamefont {{Xia}}, \citenamefont {{Duan}}, \citenamefont
  {{Shen}}, \citenamefont {{Li}},\ and\ \citenamefont {{Fan}}}]{Liang17line3}%
  \BibitemOpen
  \bibfield  {author} {\bibinfo {author} {\bibfnamefont {Y.-F.}\ \bibnamefont
  {{Liang}}}, \bibinfo {author} {\bibfnamefont {Z.-Q.}\ \bibnamefont {{Xia}}},
  \bibinfo {author} {\bibfnamefont {K.-K.}\ \bibnamefont {{Duan}}}, \bibinfo
  {author} {\bibfnamefont {Z.-Q.}\ \bibnamefont {{Shen}}}, \bibinfo {author}
  {\bibfnamefont {X.}~\bibnamefont {{Li}}}, and\ \bibinfo {author}
  {\bibfnamefont {Y.-Z.}\ \bibnamefont {{Fan}}},\ }\bibfield  {title} {\enquote
  {\bibinfo {title} {{Limits on dark matter annihilation cross sections to
  gamma-ray lines with subhalo distributions in N -body simulations and Fermi
  LAT data}},}\ }\href {\doibase 10.1103/PhysRevD.95.063531} {\bibfield
  {journal} {\bibinfo  {journal} {\prd}\ }\textbf {\bibinfo {volume} {95}},\
  \bibinfo {eid} {063531} (\bibinfo {year} {2017})},\ \Eprint
  {http://arxiv.org/abs/1703.07078}{arXiv:1703.07078}\BibitemShut {NoStop}%
\bibitem [{\citenamefont {{Feng}}\ \emph {{\it et~al.}}(2016)\citenamefont
  {{Feng}}, \citenamefont {{Liang}}, \citenamefont {{Dong}},\ and\
  \citenamefont {{Fan}}}]{fl16}%
  \BibitemOpen
  \bibfield  {author} {\bibinfo {author} {\bibfnamefont {L.}~\bibnamefont
  {{Feng}}}, \bibinfo {author} {\bibfnamefont {Y.-F.}\ \bibnamefont {{Liang}}},
  \bibinfo {author} {\bibfnamefont {T.-K.}\ \bibnamefont {{Dong}}}, and\
  \bibinfo {author} {\bibfnamefont {Y.-Z.}\ \bibnamefont {{Fan}}},\ }\bibfield
  {title} {\enquote {\bibinfo {title} {{Interpretations of the possible 42.7
  GeV {$\gamma$} -ray line}},}\ }\href {\doibase 10.1103/PhysRevD.94.043535}
  {\bibfield  {journal} {\bibinfo  {journal} {\prd}\ }\textbf {\bibinfo
  {volume} {94}},\ \bibinfo {eid} {043535} (\bibinfo {year} {2016})},\ \Eprint
  {http://arxiv.org/abs/1608.04056}{arXiv:1608.04056}\BibitemShut {NoStop}%
\bibitem [{\citenamefont {{G{\'o}rski}}\ \emph {{\it
  et~al.}}(2005)\citenamefont {{G{\'o}rski}}, \citenamefont {{Hivon}},
  \citenamefont {{Banday}}, \citenamefont {{Wandelt}}, \citenamefont
  {{Hansen}}, \citenamefont {{Reinecke}},\ and\ \citenamefont
  {{Bartelmann}}}]{healpix05}%
  \BibitemOpen
  \bibfield  {author} {\bibinfo {author} {\bibfnamefont {K.~M.}\ \bibnamefont
  {{G{\'o}rski}}}, \bibinfo {author} {\bibfnamefont {E.}~\bibnamefont
  {{Hivon}}}, \bibinfo {author} {\bibfnamefont {A.~J.}\ \bibnamefont
  {{Banday}}}, \bibinfo {author} {\bibfnamefont {B.~D.}\ \bibnamefont
  {{Wandelt}}}, \bibinfo {author} {\bibfnamefont {F.~K.}\ \bibnamefont
  {{Hansen}}}, \bibinfo {author} {\bibfnamefont {M.}~\bibnamefont
  {{Reinecke}}}, and\ \bibinfo {author} {\bibfnamefont {M.}~\bibnamefont
  {{Bartelmann}}},\ }\bibfield  {title} {\enquote {\bibinfo {title} {{HEALPix:
  A Framework for High-Resolution Discretization and Fast Analysis of Data
  Distributed on the Sphere}},}\ }\href {\doibase 10.1086/427976} {\bibfield
  {journal} {\bibinfo  {journal} {Astrophys. J.}\ }\textbf {\bibinfo {volume}
  {622}},\ \bibinfo {pages} {759} (\bibinfo {year} {2005})},\ \Eprint
  {http://arxiv.org/abs/astro-ph/0409513}{astro-ph/0409513}\BibitemShut
  {NoStop}%
\bibitem [{\citenamefont {{Ackermann}}\ \emph {{\it
  et~al.}}(2012{\natexlab{b}})\citenamefont {{Ackermann}}, \citenamefont
  {{Ajello}}, \citenamefont {{Albert}}, \citenamefont {{Allafort}},
  \citenamefont {{Atwood}}, \citenamefont {{Axelsson}}, \citenamefont
  {{Baldini}}, \citenamefont {{Ballet}}, \citenamefont {{Barbiellini}},
  \citenamefont {{Bastieri}}, \citenamefont {{Bechtol}}, \citenamefont
  {{Bellazzini}}, \citenamefont {{Bissaldi}}, \citenamefont {{Blandford}},
  \citenamefont {{Bloom}}, \citenamefont {{Bogart}}, \citenamefont
  {{Bonamente}}, \citenamefont {{Borgland}}, \citenamefont {{Bottacini}},
  \citenamefont {{Bouvier}}, \citenamefont {{Brandt}}, \citenamefont
  {{Bregeon}}, \citenamefont {{Brigida}}, \citenamefont {{Bruel}},
  \citenamefont {{Buehler}}, \citenamefont {{Burnett}}, \citenamefont
  {{Buson}}, \citenamefont {{Caliandro}}, \citenamefont {{Cameron}},
  \citenamefont {{Caraveo}}, \citenamefont {{Casandjian}}, \citenamefont
  {{Cavazzuti}}, \citenamefont {{Cecchi}}, \citenamefont {{{\c C}elik}},
  \citenamefont {{Charles}}, \citenamefont {{Chaves}}, \citenamefont
  {{Chekhtman}}, \citenamefont {{Cheung}}, \citenamefont {{Chiang}},
  \citenamefont {{Ciprini}}, \citenamefont {{Claus}}, \citenamefont
  {{Cohen-Tanugi}}, \citenamefont {{Conrad}}, \citenamefont {{Corbet}},
  \citenamefont {{Cutini}}, \citenamefont {{D'Ammando}}, \citenamefont
  {{Davis}}, \citenamefont {{de Angelis}}, \citenamefont {{DeKlotz}},
  \citenamefont {{de Palma}}, \citenamefont {{Dermer}}, \citenamefont
  {{Digel}}, \citenamefont {{Silva}}, \citenamefont {{Drell}}, \citenamefont
  {{Drlica-Wagner}}, \citenamefont {{Dubois}}, \citenamefont {{Favuzzi}},
  \citenamefont {{Fegan}}, \citenamefont {{Ferrara}}, \citenamefont {{Focke}},
  \citenamefont {{Fortin}}, \citenamefont {{Fukazawa}}, \citenamefont {{Funk}},
  \citenamefont {{Fusco}}, \citenamefont {{Gargano}}, \citenamefont
  {{Gasparrini}}, \citenamefont {{Gehrels}}, \citenamefont {{Giebels}},
  \citenamefont {{Giglietto}}, \citenamefont {{Giordano}}, \citenamefont
  {{Giroletti}}, \citenamefont {{Glanzman}}, \citenamefont {{Godfrey}},
  \citenamefont {{Grenier}}, \citenamefont {{Grove}}, \citenamefont
  {{Guiriec}}, \citenamefont {{Hadasch}}, \citenamefont {{Hayashida}},
  \citenamefont {{Hays}}, \citenamefont {{Horan}}, \citenamefont {{Hou}},
  \citenamefont {{Hughes}}, \citenamefont {{Jackson}}, \citenamefont
  {{Jogler}}, \citenamefont {{J{\'o}hannesson}}, \citenamefont {{Johnson}},
  \citenamefont {{Johnson}}, \citenamefont {{Johnson}}, \citenamefont
  {{Kamae}}, \citenamefont {{Katagiri}}, \citenamefont {{Kataoka}},
  \citenamefont {{Kerr}}, \citenamefont {{Kn{\"o}dlseder}}, \citenamefont
  {{Kuss}}, \citenamefont {{Lande}}, \citenamefont {{Larsson}}, \citenamefont
  {{Latronico}}, \citenamefont {{Lavalley}}, \citenamefont {{Lemoine-Goumard}},
  \citenamefont {{Longo}}, \citenamefont {{Loparco}}, \citenamefont {{Lott}},
  \citenamefont {{Lovellette}}, \citenamefont {{Lubrano}}, \citenamefont
  {{Mazziotta}}, \citenamefont {{McConville}}, \citenamefont {{McEnery}},
  \citenamefont {{Mehault}}, \citenamefont {{Michelson}}, \citenamefont
  {{Mitthumsiri}}, \citenamefont {{Mizuno}}, \citenamefont {{Moiseev}},
  \citenamefont {{Monte}}, \citenamefont {{Monzani}}, \citenamefont
  {{Morselli}}, \citenamefont {{Moskalenko}}, \citenamefont {{Murgia}},
  \citenamefont {{Naumann-Godo}}, \citenamefont {{Nemmen}}, \citenamefont
  {{Nishino}}, \citenamefont {{Norris}}, \citenamefont {{Nuss}}, \citenamefont
  {{Ohno}}, \citenamefont {{Ohsugi}}, \citenamefont {{Okumura}}, \citenamefont
  {{Omodei}}, \citenamefont {{Orienti}}, \citenamefont {{Orlando}},
  \citenamefont {{Ormes}}, \citenamefont {{Paneque}}, \citenamefont
  {{Panetta}}, \citenamefont {{Perkins}}, \citenamefont {{Pesce-Rollins}},
  \citenamefont {{Pierbattista}}, \citenamefont {{Piron}}, \citenamefont
  {{Pivato}}, \citenamefont {{Porter}}, \citenamefont {{Racusin}},
  \citenamefont {{Rain{\`o}}}, \citenamefont {{Rando}}, \citenamefont
  {{Razzano}}, \citenamefont {{Razzaque}}, \citenamefont {{Reimer}},
  \citenamefont {{Reimer}}, \citenamefont {{Reposeur}}, \citenamefont
  {{Reyes}}, \citenamefont {{Ritz}}, \citenamefont {{Rochester}}, \citenamefont
  {{Romoli}}, \citenamefont {{Roth}}, \citenamefont {{Sadrozinski}},
  \citenamefont {{Sanchez}}, \citenamefont {{Saz Parkinson}}, \citenamefont
  {{Sbarra}}, \citenamefont {{Scargle}}, \citenamefont {{Sgr{\`o}}},
  \citenamefont {{Siegal-Gaskins}}, \citenamefont {{Siskind}}, \citenamefont
  {{Spandre}}, \citenamefont {{Spinelli}}, \citenamefont {{Stephens}},
  \citenamefont {{Suson}}, \citenamefont {{Tajima}}, \citenamefont
  {{Takahashi}}, \citenamefont {{Tanaka}}, \citenamefont {{Thayer}},
  \citenamefont {{Thayer}}, \citenamefont {{Thompson}}, \citenamefont
  {{Tibaldo}}, \citenamefont {{Tinivella}}, \citenamefont {{Tosti}},
  \citenamefont {{Troja}}, \citenamefont {{Usher}}, \citenamefont
  {{Vandenbroucke}}, \citenamefont {{Van Klaveren}}, \citenamefont
  {{Vasileiou}}, \citenamefont {{Vianello}}, \citenamefont {{Vitale}},
  \citenamefont {{Waite}}, \citenamefont {{Wallace}}, \citenamefont {{Winer}},
  \citenamefont {{Wood}}, \citenamefont {{Wood}}, \citenamefont {{Wood}},
  \citenamefont {{Yang}},\ and\ \citenamefont {{Zimmer}}}]{fermi12cal}%
  \BibitemOpen
  \bibfield  {author} {\bibinfo {author} {\bibfnamefont {M.}~\bibnamefont
  {{Ackermann}}} {\it et~al.},\ }\bibfield  {title} {\enquote {\bibinfo {title}
  {{The Fermi Large Area Telescope on Orbit: Event Classification, Instrument
  Response Functions, and Calibration}},}\ }\href {\doibase
  10.1088/0067-0049/203/1/4} {\bibfield  {journal} {\bibinfo  {journal}
  {Astrophys. J. Suppl.}\ }\textbf {\bibinfo {volume} {203}},\ \bibinfo {eid}
  {4} (\bibinfo {year} {2012}{\natexlab{b}})},\ \Eprint
  {http://arxiv.org/abs/1206.1896}{arXiv:1206.1896}\BibitemShut {NoStop}%
\bibitem [{\citenamefont {{Lefranc}}\ \emph {{\it et~al.}}(2016)\citenamefont
  {{Lefranc}}, \citenamefont {{Moulin}}, \citenamefont {{Panci}}, \citenamefont
  {{Sala}},\ and\ \citenamefont {{Silk}}}]{Lefranc:2016fgn}%
  \BibitemOpen
  \bibfield  {author} {\bibinfo {author} {\bibfnamefont {V.}~\bibnamefont
  {{Lefranc}}}, \bibinfo {author} {\bibfnamefont {E.}~\bibnamefont {{Moulin}}},
  \bibinfo {author} {\bibfnamefont {P.}~\bibnamefont {{Panci}}}, \bibinfo
  {author} {\bibfnamefont {F.}~\bibnamefont {{Sala}}}, and\ \bibinfo {author}
  {\bibfnamefont {J.}~\bibnamefont {{Silk}}},\ }\bibfield  {title} {\enquote
  {\bibinfo {title} {{Dark Matter in {$\gamma$} lines: Galactic Center vs.
  dwarf galaxies}},}\ }\href {\doibase 10.1088/1475-7516/2016/09/043}
  {\bibfield  {journal} {\bibinfo  {journal} {\jcap}\ }\textbf {\bibinfo
  {volume} {9}},\ \bibinfo {eid} {043} (\bibinfo {year} {2016})},\ \Eprint
  {http://arxiv.org/abs/1608.00786}{arXiv:1608.00786}\BibitemShut {NoStop}%
\bibitem [{\citenamefont {{Diemand}}\ \emph {{\it et~al.}}(2008)\citenamefont
  {{Diemand}}, \citenamefont {{Kuhlen}}, \citenamefont {{Madau}}, \citenamefont
  {{Zemp}}, \citenamefont {{Moore}}, \citenamefont {{Potter}},\ and\
  \citenamefont {{Stadel}}}]{diemand08VL2}%
  \BibitemOpen
  \bibfield  {author} {\bibinfo {author} {\bibfnamefont {J.}~\bibnamefont
  {{Diemand}}}, \bibinfo {author} {\bibfnamefont {M.}~\bibnamefont {{Kuhlen}}},
  \bibinfo {author} {\bibfnamefont {P.}~\bibnamefont {{Madau}}}, \bibinfo
  {author} {\bibfnamefont {M.}~\bibnamefont {{Zemp}}}, \bibinfo {author}
  {\bibfnamefont {B.}~\bibnamefont {{Moore}}}, \bibinfo {author} {\bibfnamefont
  {D.}~\bibnamefont {{Potter}}}, and\ \bibinfo {author} {\bibfnamefont
  {J.}~\bibnamefont {{Stadel}}},\ }\bibfield  {title} {\enquote {\bibinfo
  {title} {{Clumps and streams in the local dark matter distribution}},}\
  }\href {\doibase 10.1038/nature07153} {\bibfield  {journal} {\bibinfo
  {journal} {Nature}\ }\textbf {\bibinfo {volume} {454}},\ \bibinfo {pages}
  {735} (\bibinfo {year} {2008})},\ \Eprint
  {http://arxiv.org/abs/0805.1244}{arXiv:0805.1244}\BibitemShut {NoStop}%
\bibitem [{\citenamefont {{Springel}}\ \emph {{\it et~al.}}(2008)\citenamefont
  {{Springel}}, \citenamefont {{Wang}}, \citenamefont {{Vogelsberger}},
  \citenamefont {{Ludlow}}, \citenamefont {{Jenkins}}, \citenamefont {{Helmi}},
  \citenamefont {{Navarro}}, \citenamefont {{Frenk}},\ and\ \citenamefont
  {{White}}}]{springel08Aquarius}%
  \BibitemOpen
  \bibfield  {author} {\bibinfo {author} {\bibfnamefont {V.}~\bibnamefont
  {{Springel}}}, \bibinfo {author} {\bibfnamefont {J.}~\bibnamefont {{Wang}}},
  \bibinfo {author} {\bibfnamefont {M.}~\bibnamefont {{Vogelsberger}}},
  \bibinfo {author} {\bibfnamefont {A.}~\bibnamefont {{Ludlow}}}, \bibinfo
  {author} {\bibfnamefont {A.}~\bibnamefont {{Jenkins}}}, \bibinfo {author}
  {\bibfnamefont {A.}~\bibnamefont {{Helmi}}}, \bibinfo {author} {\bibfnamefont
  {J.~F.}\ \bibnamefont {{Navarro}}}, \bibinfo {author} {\bibfnamefont {C.~S.}\
  \bibnamefont {{Frenk}}}, and\ \bibinfo {author} {\bibfnamefont {S.~D.~M.}\
  \bibnamefont {{White}}},\ }\bibfield  {title} {\enquote {\bibinfo {title}
  {{The Aquarius Project: the subhaloes of galactic haloes}},}\ }\href
  {\doibase 10.1111/j.1365-2966.2008.14066.x} {\bibfield  {journal} {\bibinfo
  {journal} {\mnras}\ }\textbf {\bibinfo {volume} {391}},\ \bibinfo {pages}
  {1685} (\bibinfo {year} {2008})},\ \Eprint
  {http://arxiv.org/abs/0809.0898}{arXiv:0809.0898}\BibitemShut {NoStop}%
\bibitem [{\citenamefont {{Garrison-Kimmel}}\ \emph {{\it
  et~al.}}(2014)\citenamefont {{Garrison-Kimmel}}, \citenamefont
  {{Boylan-Kolchin}}, \citenamefont {{Bullock}},\ and\ \citenamefont
  {{Lee}}}]{garrison14ELVIS}%
  \BibitemOpen
  \bibfield  {author} {\bibinfo {author} {\bibfnamefont {S.}~\bibnamefont
  {{Garrison-Kimmel}}}, \bibinfo {author} {\bibfnamefont {M.}~\bibnamefont
  {{Boylan-Kolchin}}}, \bibinfo {author} {\bibfnamefont {J.~S.}\ \bibnamefont
  {{Bullock}}}, and\ \bibinfo {author} {\bibfnamefont {K.}~\bibnamefont
  {{Lee}}},\ }\bibfield  {title} {\enquote {\bibinfo {title} {{ELVIS: Exploring
  the Local Volume in Simulations}},}\ }\href {\doibase 10.1093/mnras/stt2377}
  {\bibfield  {journal} {\bibinfo  {journal} {\mnras}\ }\textbf {\bibinfo
  {volume} {438}},\ \bibinfo {pages} {2578} (\bibinfo {year} {2014})},\ \Eprint
  {http://arxiv.org/abs/1310.6746}{arXiv:1310.6746}\BibitemShut {NoStop}%
\bibitem [{\citenamefont {{Ackermann}}\ \emph {{\it
  et~al.}}(2012{\natexlab{c}})\citenamefont {{Ackermann}} {\it
  et~al.}}]{fermi12dmsh}%
  \BibitemOpen
  \bibfield  {author} {\bibinfo {author} {\bibfnamefont {M.}~\bibnamefont
  {{Ackermann}}}  {\it et~al.} (\bibinfo {collaboration} {Fermi LAT}
  Collaboration),\ }\bibfield  {title} {\enquote {\bibinfo {title} {{Search for
  Dark Matter Satellites Using Fermi-LAT}},}\ }\href {\doibase
  10.1088/0004-637X/747/2/121} {\bibfield  {journal} {\bibinfo  {journal}
  {Astrophys. J.}\ }\textbf {\bibinfo {volume} {747}},\ \bibinfo {eid} {121}
  (\bibinfo {year} {2012}{\natexlab{c}})},\ \Eprint
  {http://arxiv.org/abs/1201.2691}{arXiv:1201.2691}\BibitemShut {NoStop}%
\bibitem [{\citenamefont {{Berlin}}\ and\ \citenamefont
  {{Hooper}}(2014)}]{berlin14}%
  \BibitemOpen
  \bibfield  {author} {\bibinfo {author} {\bibfnamefont {A.}~\bibnamefont
  {{Berlin}}} and\ \bibinfo {author} {\bibfnamefont {D.}~\bibnamefont
  {{Hooper}}},\ }\bibfield  {title} {\enquote {\bibinfo {title} {{Stringent
  constraints on the dark matter annihilation cross section from subhalo
  searches with the Fermi Gamma-Ray Space Telescope}},}\ }\href {\doibase
  10.1103/PhysRevD.89.016014} {\bibfield  {journal} {\bibinfo  {journal}
  {\prd}\ }\textbf {\bibinfo {volume} {89}},\ \bibinfo {eid} {016014} (\bibinfo
  {year} {2014})},\ \Eprint
  {http://arxiv.org/abs/1309.0525}{arXiv:1309.0525}\BibitemShut {NoStop}%
\bibitem [{\citenamefont {{Bertoni}}\ \emph {{\it et~al.}}(2015)\citenamefont
  {{Bertoni}}, \citenamefont {{Hooper}},\ and\ \citenamefont
  {{Linden}}}]{bertoni15dmsh}%
  \BibitemOpen
  \bibfield  {author} {\bibinfo {author} {\bibfnamefont {B.}~\bibnamefont
  {{Bertoni}}}, \bibinfo {author} {\bibfnamefont {D.}~\bibnamefont {{Hooper}}},
  and\ \bibinfo {author} {\bibfnamefont {T.}~\bibnamefont {{Linden}}},\
  }\bibfield  {title} {\enquote {\bibinfo {title} {{Examining The Fermi-LAT
  Third Source Catalog in search of dark matter subhalos}},}\ }\href {\doibase
  10.1088/1475-7516/2015/12/035} {\bibfield  {journal} {\bibinfo  {journal}
  {\jcap}\ }\textbf {\bibinfo {volume} {12}},\ \bibinfo {eid} {035} (\bibinfo
  {year} {2015})},\ \Eprint
  {http://arxiv.org/abs/1504.02087}{arXiv:1504.02087}\BibitemShut {NoStop}%
\bibitem [{\citenamefont {{Schoonenberg}}\ \emph {{\it
  et~al.}}(2016)\citenamefont {{Schoonenberg}}, \citenamefont {{Gaskins}},
  \citenamefont {{Bertone}},\ and\ \citenamefont
  {{Diemand}}}]{schoonenberg16dmsh}%
  \BibitemOpen
  \bibfield  {author} {\bibinfo {author} {\bibfnamefont {D.}~\bibnamefont
  {{Schoonenberg}}}, \bibinfo {author} {\bibfnamefont {J.}~\bibnamefont
  {{Gaskins}}}, \bibinfo {author} {\bibfnamefont {G.}~\bibnamefont
  {{Bertone}}}, and\ \bibinfo {author} {\bibfnamefont {J.}~\bibnamefont
  {{Diemand}}},\ }\bibfield  {title} {\enquote {\bibinfo {title} {{Dark matter
  subhalos and unidentified sources in the Fermi 3FGL source catalog}},}\
  }\href {\doibase 10.1088/1475-7516/2016/05/028} {\bibfield  {journal}
  {\bibinfo  {journal} {\jcap}\ }\textbf {\bibinfo {volume} {5}},\ \bibinfo
  {eid} {028} (\bibinfo {year} {2016})},\ \Eprint
  {http://arxiv.org/abs/1601.06781}{arXiv:1601.06781}\BibitemShut {NoStop}%
\bibitem [{\citenamefont {{Bertoni}}\ \emph {{\it et~al.}}(2016)\citenamefont
  {{Bertoni}}, \citenamefont {{Hooper}},\ and\ \citenamefont
  {{Linden}}}]{bertoni16j2212}%
  \BibitemOpen
  \bibfield  {author} {\bibinfo {author} {\bibfnamefont {B.}~\bibnamefont
  {{Bertoni}}}, \bibinfo {author} {\bibfnamefont {D.}~\bibnamefont {{Hooper}}},
  and\ \bibinfo {author} {\bibfnamefont {T.}~\bibnamefont {{Linden}}},\
  }\bibfield  {title} {\enquote {\bibinfo {title} {{Is the gamma-ray source
  3FGL J2212.5+0703 a dark matter subhalo?}}}\ }\href {\doibase
  10.1088/1475-7516/2016/05/049} {\bibfield  {journal} {\bibinfo  {journal}
  {\jcap}\ }\textbf {\bibinfo {volume} {5}},\ \bibinfo {eid} {049} (\bibinfo
  {year} {2016})},\ \Eprint
  {http://arxiv.org/abs/1602.07303}{arXiv:1602.07303}\BibitemShut {NoStop}%
\bibitem [{\citenamefont {{Hooper}}\ and\ \citenamefont
  {{Witte}}(2017)}]{hooper16dmsh}%
  \BibitemOpen
  \bibfield  {author} {\bibinfo {author} {\bibfnamefont {D.}~\bibnamefont
  {{Hooper}}} and\ \bibinfo {author} {\bibfnamefont {S.~J.}\ \bibnamefont
  {{Witte}}},\ }\bibfield  {title} {\enquote {\bibinfo {title} {{Gamma rays
  from dark matter subhalos revisited: refining the predictions and
  constraints}},}\ }\href {\doibase 10.1088/1475-7516/2017/04/018} {\bibfield
  {journal} {\bibinfo  {journal} {\jcap}\ }\textbf {\bibinfo {volume} {4}},\
  \bibinfo {eid} {018} (\bibinfo {year} {2017})},\ \Eprint
  {http://arxiv.org/abs/1610.07587}{arXiv:1610.07587}\BibitemShut {NoStop}%
\bibitem [{\citenamefont {{Calore}}\ \emph {{\it et~al.}}(2017)\citenamefont
  {{Calore}}, \citenamefont {{De Romeri}}, \citenamefont {{Di Mauro}},
  \citenamefont {{Donato}},\ and\ \citenamefont {{Marinacci}}}]{calore16dmsh}%
  \BibitemOpen
  \bibfield  {author} {\bibinfo {author} {\bibfnamefont {F.}~\bibnamefont
  {{Calore}}}, \bibinfo {author} {\bibfnamefont {V.}~\bibnamefont {{De
  Romeri}}}, \bibinfo {author} {\bibfnamefont {M.}~\bibnamefont {{Di Mauro}}},
  \bibinfo {author} {\bibfnamefont {F.}~\bibnamefont {{Donato}}}, and\ \bibinfo
  {author} {\bibfnamefont {F.}~\bibnamefont {{Marinacci}}},\ }\bibfield
  {title} {\enquote {\bibinfo {title} {{Realistic estimation for the
  detectability of dark matter subhalos using Fermi-LAT catalogs}},}\ }\href
  {\doibase 10.1103/PhysRevD.96.063009} {\bibfield  {journal} {\bibinfo
  {journal} {\prd}\ }\textbf {\bibinfo {volume} {96}},\ \bibinfo {eid} {063009}
  (\bibinfo {year} {2017})},\ \Eprint
  {http://arxiv.org/abs/1611.03503}{arXiv:1611.03503}\BibitemShut {NoStop}%
\bibitem [{\citenamefont {{Wang}}\ \emph {{\it et~al.}}(2016)\citenamefont
  {{Wang}}, \citenamefont {{Duan}}, \citenamefont {{Ma}}, \citenamefont
  {{Liang}}, \citenamefont {{Shen}}, \citenamefont {{Li}}, \citenamefont
  {{Yue}}, \citenamefont {{Yuan}}, \citenamefont {{Zang}}, \citenamefont
  {{Fan}},\ and\ \citenamefont {{Chang}}}]{wyp16dmsh}%
  \BibitemOpen
  \bibfield  {author} {\bibinfo {author} {\bibfnamefont {Y.-P.}\ \bibnamefont
  {{Wang}}} {\it et~al.},\ }\bibfield  {title} {\enquote {\bibinfo {title}
  {{Testing the dark matter subhalo hypothesis of the gamma-ray source 3FGL
  J2212.5 +0703}},}\ }\href {\doibase 10.1103/PhysRevD.94.123002} {\bibfield
  {journal} {\bibinfo  {journal} {\prd}\ }\textbf {\bibinfo {volume} {94}},\
  \bibinfo {eid} {123002} (\bibinfo {year} {2016})},\ \Eprint
  {http://arxiv.org/abs/1611.05135}{arXiv:1611.05135}\BibitemShut {NoStop}%
\bibitem [{\citenamefont {{Xia}}\ \emph {{\it et~al.}}(2017)\citenamefont
  {{Xia}}, \citenamefont {{Duan}}, \citenamefont {{Li}}, \citenamefont
  {{Liang}}, \citenamefont {{Shen}}, \citenamefont {{Yue}}, \citenamefont
  {{Wang}}, \citenamefont {{Yuan}}, \citenamefont {{Fan}}, \citenamefont
  {{Wu}},\ and\ \citenamefont {{Chang}}}]{xzq16dmsh}%
  \BibitemOpen
  \bibfield  {author} {\bibinfo {author} {\bibfnamefont {Z.-Q.}\ \bibnamefont
  {{Xia}}} {\it et~al.},\ }\bibfield  {title} {\enquote {\bibinfo {title}
  {{3FGL J1924.8-1034: A spatially extended stable unidentified GeV source?}}}\
  }\href {\doibase 10.1103/PhysRevD.95.102001} {\bibfield  {journal} {\bibinfo
  {journal} {\prd}\ }\textbf {\bibinfo {volume} {95}},\ \bibinfo {eid} {102001}
  (\bibinfo {year} {2017})},\ \Eprint
  {http://arxiv.org/abs/1611.05565}{arXiv:1611.05565}\BibitemShut {NoStop}%
\bibitem [{\citenamefont {{Navarro}}\ \emph {{\it et~al.}}(1997)\citenamefont
  {{Navarro}}, \citenamefont {{Frenk}},\ and\ \citenamefont
  {{White}}}]{navarro97nfw}%
  \BibitemOpen
  \bibfield  {author} {\bibinfo {author} {\bibfnamefont {J.~F.}\ \bibnamefont
  {{Navarro}}}, \bibinfo {author} {\bibfnamefont {C.~S.}\ \bibnamefont
  {{Frenk}}}, and\ \bibinfo {author} {\bibfnamefont {S.~D.~M.}\ \bibnamefont
  {{White}}},\ }\bibfield  {title} {\enquote {\bibinfo {title} {{A Universal
  Density Profile from Hierarchical Clustering}},}\ }\href {\doibase
  10.1086/304888} {\bibfield  {journal} {\bibinfo  {journal} {Astrophys. J.}\
  }\textbf {\bibinfo {volume} {490}},\ \bibinfo {pages} {493} (\bibinfo {year}
  {1997})},\ \Eprint
  {http://arxiv.org/abs/astro-ph/9611107}{astro-ph/9611107}\BibitemShut
  {NoStop}%
\bibitem [{\citenamefont {{Acero}}\ \emph {{\it et~al.}}(2015)\citenamefont
  {{Acero}}, \citenamefont {{Ackermann}}, \citenamefont {{Ajello}},
  \citenamefont {{Albert}}, \citenamefont {{Atwood}}, \citenamefont
  {{Axelsson}}, \citenamefont {{Baldini}}, \citenamefont {{Ballet}},
  \citenamefont {{Barbiellini}}, \citenamefont {{Bastieri}}, \citenamefont
  {{Belfiore}}, \citenamefont {{Bellazzini}}, \citenamefont {{Bissaldi}},
  \citenamefont {{Blandford}}, \citenamefont {{Bloom}}, \citenamefont
  {{Bogart}}, \citenamefont {{Bonino}}, \citenamefont {{Bottacini}},
  \citenamefont {{Bregeon}}, \citenamefont {{Britto}}, \citenamefont {{Bruel}},
  \citenamefont {{Buehler}}, \citenamefont {{Burnett}}, \citenamefont
  {{Buson}}, \citenamefont {{Caliandro}}, \citenamefont {{Cameron}},
  \citenamefont {{Caputo}}, \citenamefont {{Caragiulo}}, \citenamefont
  {{Caraveo}}, \citenamefont {{Casandjian}}, \citenamefont {{Cavazzuti}},
  \citenamefont {{Charles}}, \citenamefont {{Chaves}}, \citenamefont
  {{Chekhtman}}, \citenamefont {{Cheung}}, \citenamefont {{Chiang}},
  \citenamefont {{Chiaro}}, \citenamefont {{Ciprini}}, \citenamefont {{Claus}},
  \citenamefont {{Cohen-Tanugi}}, \citenamefont {{Cominsky}}, \citenamefont
  {{Conrad}}, \citenamefont {{Cutini}}, \citenamefont {{D'Ammando}},
  \citenamefont {{de Angelis}}, \citenamefont {{DeKlotz}}, \citenamefont {{de
  Palma}}, \citenamefont {{Desiante}}, \citenamefont {{Digel}}, \citenamefont
  {{Di Venere}}, \citenamefont {{Drell}}, \citenamefont {{Dubois}},
  \citenamefont {{Dumora}}, \citenamefont {{Favuzzi}}, \citenamefont {{Fegan}},
  \citenamefont {{Ferrara}}, \citenamefont {{Finke}}, \citenamefont
  {{Franckowiak}}, \citenamefont {{Fukazawa}}, \citenamefont {{Funk}},
  \citenamefont {{Fusco}}, \citenamefont {{Gargano}}, \citenamefont
  {{Gasparrini}}, \citenamefont {{Giebels}}, \citenamefont {{Giglietto}},
  \citenamefont {{Giommi}}, \citenamefont {{Giordano}}, \citenamefont
  {{Giroletti}}, \citenamefont {{Glanzman}}, \citenamefont {{Godfrey}},
  \citenamefont {{Grenier}}, \citenamefont {{Grondin}}, \citenamefont
  {{Grove}}, \citenamefont {{Guillemot}}, \citenamefont {{Guiriec}},
  \citenamefont {{Hadasch}}, \citenamefont {{Harding}}, \citenamefont {{Hays}},
  \citenamefont {{Hewitt}}, \citenamefont {{Hill}}, \citenamefont {{Horan}},
  \citenamefont {{Iafrate}}, \citenamefont {{Jogler}}, \citenamefont
  {{J{\'o}hannesson}}, \citenamefont {{Johnson}}, \citenamefont {{Johnson}},
  \citenamefont {{Johnson}}, \citenamefont {{Johnson}}, \citenamefont
  {{Kamae}}, \citenamefont {{Kataoka}}, \citenamefont {{Katsuta}},
  \citenamefont {{Kuss}}, \citenamefont {{La Mura}}, \citenamefont {{Landriu}},
  \citenamefont {{Larsson}}, \citenamefont {{Latronico}}, \citenamefont
  {{Lemoine-Goumard}}, \citenamefont {{Li}}, \citenamefont {{Li}},
  \citenamefont {{Longo}}, \citenamefont {{Loparco}}, \citenamefont {{Lott}},
  \citenamefont {{Lovellette}}, \citenamefont {{Lubrano}}, \citenamefont
  {{Madejski}}, \citenamefont {{Massaro}}, \citenamefont {{Mayer}},
  \citenamefont {{Mazziotta}}, \citenamefont {{McEnery}}, \citenamefont
  {{Michelson}}, \citenamefont {{Mirabal}}, \citenamefont {{Mizuno}},
  \citenamefont {{Moiseev}}, \citenamefont {{Mongelli}}, \citenamefont
  {{Monzani}}, \citenamefont {{Morselli}}, \citenamefont {{Moskalenko}},
  \citenamefont {{Murgia}}, \citenamefont {{Nuss}}, \citenamefont {{Ohno}},
  \citenamefont {{Ohsugi}}, \citenamefont {{Omodei}}, \citenamefont
  {{Orienti}}, \citenamefont {{Orlando}}, \citenamefont {{Ormes}},
  \citenamefont {{Paneque}}, \citenamefont {{Panetta}}, \citenamefont
  {{Perkins}}, \citenamefont {{Pesce-Rollins}}, \citenamefont {{Piron}},
  \citenamefont {{Pivato}}, \citenamefont {{Porter}}, \citenamefont
  {{Racusin}}, \citenamefont {{Rando}}, \citenamefont {{Razzano}},
  \citenamefont {{Razzaque}}, \citenamefont {{Reimer}}, \citenamefont
  {{Reimer}}, \citenamefont {{Reposeur}}, \citenamefont {{Rochester}},
  \citenamefont {{Romani}}, \citenamefont {{Salvetti}}, \citenamefont
  {{S{\'a}nchez-Conde}}, \citenamefont {{Saz Parkinson}}, \citenamefont
  {{Schulz}}, \citenamefont {{Siskind}}, \citenamefont {{Smith}}, \citenamefont
  {{Spada}}, \citenamefont {{Spandre}}, \citenamefont {{Spinelli}},
  \citenamefont {{Stephens}}, \citenamefont {{Strong}}, \citenamefont
  {{Suson}}, \citenamefont {{Takahashi}}, \citenamefont {{Takahashi}},
  \citenamefont {{Tanaka}}, \citenamefont {{Thayer}}, \citenamefont {{Thayer}},
  \citenamefont {{Thompson}}, \citenamefont {{Tibaldo}}, \citenamefont
  {{Tibolla}}, \citenamefont {{Torres}}, \citenamefont {{Torresi}},
  \citenamefont {{Tosti}}, \citenamefont {{Troja}}, \citenamefont {{Van
  Klaveren}}, \citenamefont {{Vianello}}, \citenamefont {{Winer}},
  \citenamefont {{Wood}}, \citenamefont {{Wood}}, \citenamefont {{Zimmer}},\
  and\ \citenamefont {{Fermi-LAT Collaboration}}}]{fermi15_3fgl}%
  \BibitemOpen
  \bibfield  {author} {\bibinfo {author} {\bibfnamefont {F.}~\bibnamefont
  {{Acero}}} {\it et~al.} (\bibinfo {collaboration} {Fermi LAT}
  Collaboration),\ }\bibfield  {title} {\enquote {\bibinfo {title} {{Fermi
  Large Area Telescope Third Source Catalog}},}\ }\href {\doibase
  10.1088/0067-0049/218/2/23} {\bibfield  {journal} {\bibinfo  {journal}
  {Astrophys. J. Suppl.}\ }\textbf {\bibinfo {volume} {218}},\ \bibinfo {eid}
  {23} (\bibinfo {year} {2015})},\ \Eprint
  {http://arxiv.org/abs/1501.02003}{arXiv:1501.02003}\BibitemShut {NoStop}%
\bibitem [{\citenamefont {{Galper}}\ \emph {{\it et~al.}}(2014)\citenamefont
  {{Galper}} {\it et~al.}}]{galper14gamma400}%
  \BibitemOpen
  \bibfield  {author} {\bibinfo {author} {\bibfnamefont {A.~M.}\ \bibnamefont
  {{Galper}}}  {\it et~al.},\ }\bibfield  {title} {\enquote {\bibinfo {title}
  {{The GAMMA-400 space observatory: status and perspectives}},}\ }\href@noop
  {} {\bibfield  {journal} {\bibinfo  {journal} {ArXiv e-prints}\ } (\bibinfo
  {year} {2014})},\ \Eprint
  {http://arxiv.org/abs/1412.4239}{arXiv:1412.4239}\BibitemShut {NoStop}%
\bibitem [{\citenamefont {{Zhang}}\ \emph {{\it et~al.}}(2014)\citenamefont
  {{Zhang}} {\it et~al.}}]{zhang14herd}%
  \BibitemOpen
  \bibfield  {author} {\bibinfo {author} {\bibfnamefont {S.~N.}\ \bibnamefont
  {{Zhang}}}  {\it et~al.} (\bibinfo {collaboration} {HERD} Collaboration),\
  }\bibfield  {title} {\enquote {\bibinfo {title} {{The high energy
  cosmic-radiation detection (HERD) facility onboard China's Space Station}},}\
  }\href {\doibase 10.1117/12.2055280} {\bibfield  {journal} {\bibinfo
  {journal} {Proc. SPIE Int. Soc. Opt. Eng.}\ }\textbf {\bibinfo {volume}
  {9144}},\ \bibinfo {eid} {91440X} (\bibinfo {year} {2014})},\ \Eprint
  {http://arxiv.org/abs/1407.4866}{arXiv:1407.4866}\BibitemShut {NoStop}%
\bibitem [{\citenamefont {Chernoff}(1954)}]{chernoff1954}%
  \BibitemOpen
  \bibfield  {author} {\bibinfo {author} {\bibfnamefont {H.}~\bibnamefont
  {Chernoff}},\ }\bibfield  {title} {\enquote {\bibinfo {title} {On the
  distribution of the likelihood ratio},}\ }\href {\doibase
  10.1214/aoms/1177728725} {\bibfield  {journal} {\bibinfo  {journal} {Ann.
  Math. Statist.}\ }\textbf {\bibinfo {volume} {25}},\ \bibinfo {pages} {573}
  (\bibinfo {year} {1954})}\BibitemShut {NoStop}%
\end{thebibliography}%

\appendix

\begin{figure*}[h]
\includegraphics[width=0.45\textwidth]{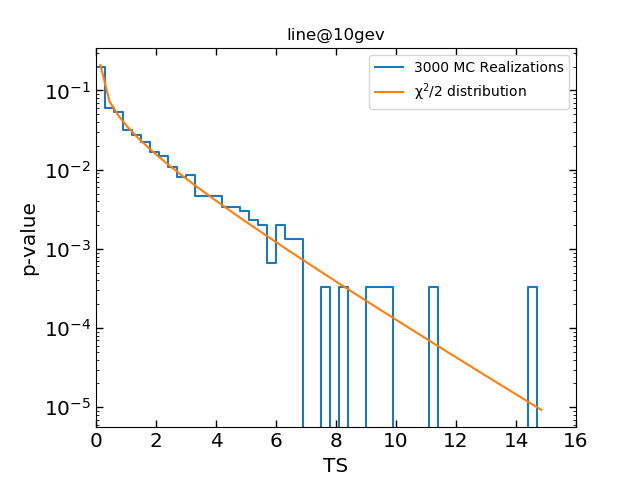}
\includegraphics[width=0.45\textwidth]{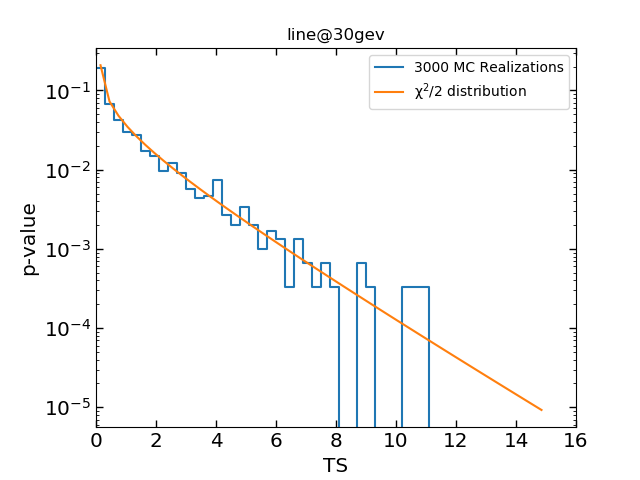}
\includegraphics[width=0.45\textwidth]{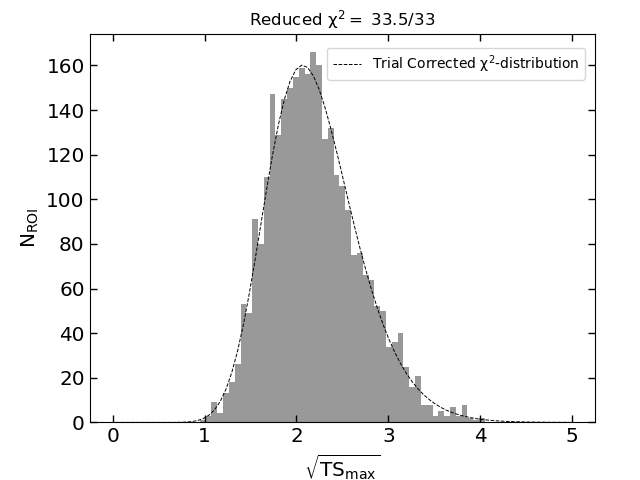}
\includegraphics[width=0.45\textwidth]{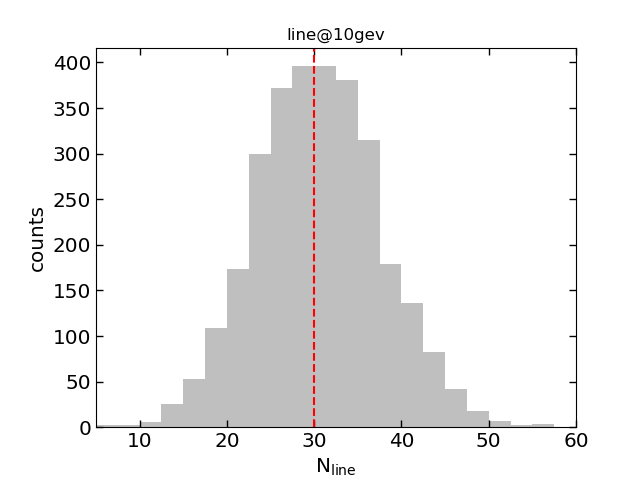}
\caption{{\tt Top:} Distribution of TS values from individual fits of a 10 GeV (left panel) or 30 GeV (right panel) line to the 3000 background-only simulations. {\tt Bottom left:} The $TS_{\rm max}$ distribution of 3000 Monte Carlo spectra generated from background-only simulations. The dashed line is the best fit trial-corrected $\chi^2$ distribution. {\tt Bottom right:} We inject a pesudo line signal with photon number $N_{\rm line}=30$ at 10 GeV into the MC data and fit them with background+line model. This plot shows that our fits can well recover the input parameter (red line).}
\label{fig:validity}
\end{figure*}

\section{Subhalo distribution and the distributions of $\gamma$ and $R_b$}
\label{distribution}

The subhalo distribution and the distributions of $\gamma$ and $R_b$ adopted in our analysis are from Ref. \cite{hooper16dmsh}.
When deriving the distributions, their dependences on both the subhalo mass and the location relative to the galactic center have been taken into account \cite{hooper16dmsh}. The subhalo distribution is
\begin{equation}\label{eq:numdens}
\frac{dN}{dM dV} = \frac{628}{M_\odot \, {\rm kpc}^3}\left(\frac{M}{M_\odot}\right)^{-1.9} \, .
\end{equation}
The distributions of $R_b$ is
\begin{eqnarray}
\frac{dP}{dR_b} &=& \frac{1}{\sigma \sqrt{2\pi}} \, \frac{1}{R_b}\exp\left(-\frac{(\ln R_b - \ln \langle R_b \rangle)^2}{2 \sigma^2} \right),
\label{probrb}
\end{eqnarray}
where $\sigma=0.496$ and $\langle R_b \rangle=10^{-3.945}\times(M/M_\odot)^{0.421}$.
The distributions of $\gamma$ is found to be independent of subhalo mass and is expressed as
\begin{eqnarray}
\frac{dP}{d\gamma} &=& \frac{1}{\sqrt{2\pi}}\,\frac{1}{\sigma-\kappa(\gamma-\langle  \gamma \rangle)} \times \nonumber\\
&&\exp\left(-\frac{\ln^2( 1 - \kappa (\gamma - \langle  \gamma \rangle) / \sigma)}{2 \kappa^2} \right),
\label{probgamma}
\end{eqnarray}
with $\langle  \gamma \rangle=0.74$, $\sigma=0.42$ and $\kappa=0.10$.

\section{Validity of the search method}
\label{app:test}

In this section we carry out some Monte Carlo (MC) simulation to test the validity of the method to search for the line (sliding window technique, unbinned analysis, etc.).

We first simulate background-only data, and fit them with the background+line model to examine the null distribution.
We model the Fermi-LAT observation spectrum averaged over all the sky (excluding the Galactic plane and the regions around bright gamma-ray sources, see Section \ref{sec:limits}) with a PLE function and use this PLE spectrum to approximate the backgrounds in our line searches. Based on this PLE background spectrum, we generate pseudo photons in 2$^\circ$ ROI according to Poisson statistics.  We apply the same search procedure as that in Sec. \ref{sec2} on these pseudo data, and derive corresponding TS values of the line component.
As predicted by the asymptotic theorem of Chernoff \cite{chernoff1954}, for individual $E_\gamma$, the distribution of the TS values (null distribution) should follow a $\chi^2/2$ distribution.
From the top panels of Figure \ref{fig:validity}, we find that the TS distribution derived from 3000 MC simulations is well consistent with matched to the theoretical prediction for two representative line energies of 10 GeV and 30 GeV. 
Applying sliding window fit to the 3000 MC spectra and deriving the $TS_{\rm max}$ of them, we obtain the distribution shown in the bottom-left panel of Figure \ref{fig:validity}. Unlike significant deviation around $TS_{\rm max}\sim4-5$ displaying in Figure \ref{fig:tsdistribution}, the trial-corrected $\chi^2$ distribution can lead to a good fit (reduced-$\chi^2=33.5/33$) to the $TS_{\rm max}$ distribution of simulation data.

Furthermore, we inject a line component with intensity of 30 photons, of which the profile is the energy dispersion function of the Fermi-LAT data used in this work, into the background to generate MC events containing a fake line signal. 
We apply the search analysis to these events to re-derive the photon number / flux of the line component.
We find that the input parameter of $N_{\rm line}=30$ can be well recovered (bottom-right panel of Figure \ref{fig:validity}).

The results presented in this section indicate our search method/code are valid to search for a line signal from the background and the deviation around $TS_{\rm max}\sim4-5$ appeared in Figure \ref{fig:tsdistribution} is most-likely irrelevant to the analysis procedure.

\section{The spectrum of the ROI giving highest TS value}
\label{spectrum}
\begin{figure}[!h]
\includegraphics[width=0.45\textwidth]{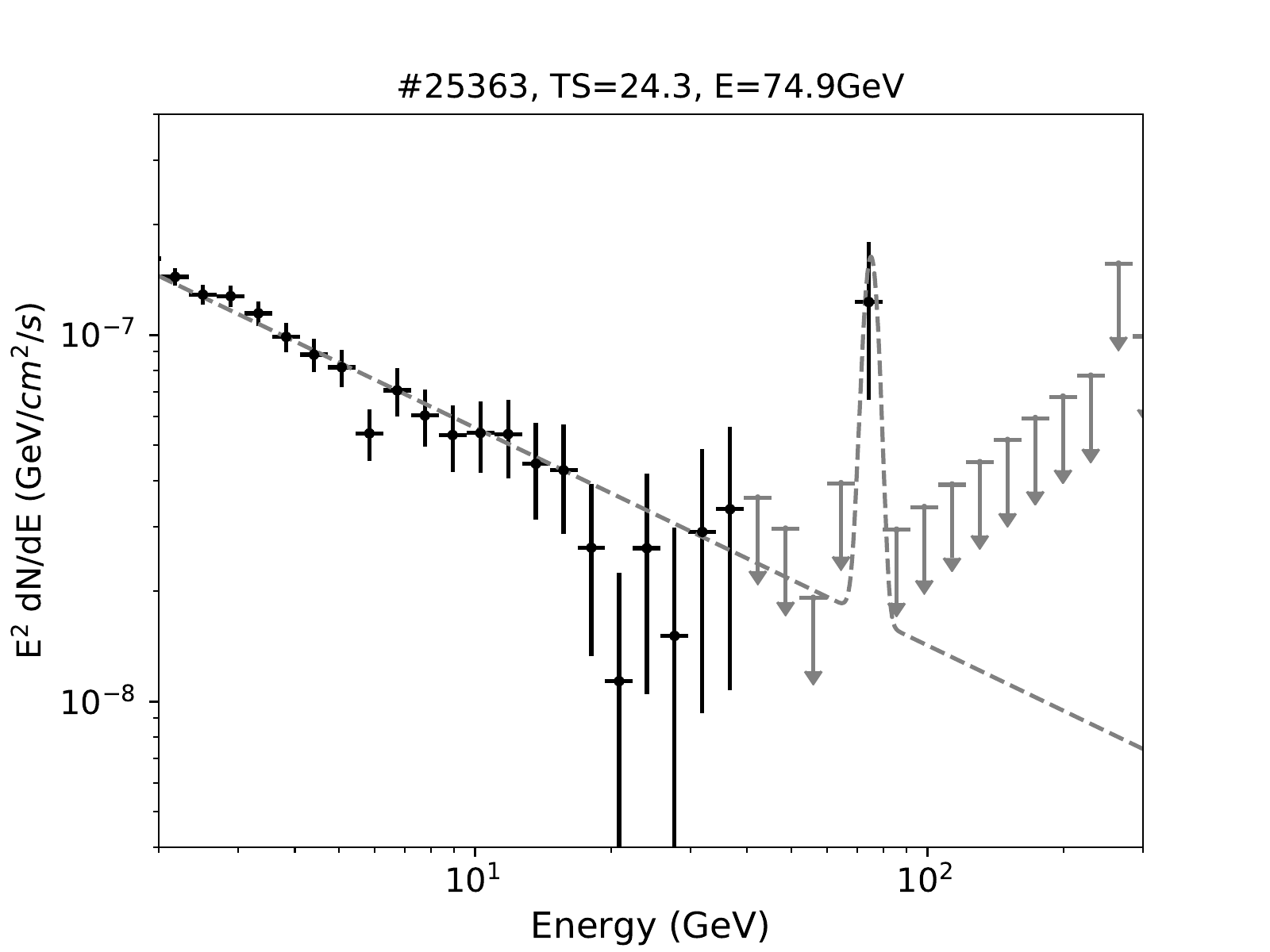}
\caption{The observed spectrum of the ROI displaying line-like excess with $TS=24.3$, which is the highest one in our analysis. For the energy bins with photon number less than 2, we plot $1\sigma$ confidence level upper limits for better visualization. Note that the line-like excess shown here is most likely from background fluctuation since a very large trial factor is introduced in the analysis.}
\label{fig:spectrum}
\end{figure}

Figure \ref{fig:spectrum} presents the Fermi-LAT spectrum of the ROI giving the highest $TS_{\rm max}$ in our analysis. 
The excess is seen around energy of 74.9 GeV, indicating that our analysis can effectively identify such a kind of signal in the spectrum.
We would like to emphasize that the binned plots in Figure \ref{fig:spectrum} are just for illustration, while in the search procedure of the main text an unbinned method is adopted.

\section{Information of 50 ROIs with $TS_{\rm max}>16$}
\label{list}
As discussed in the main text, we think the 50 ROIs with relative high significances of $TS_{\rm max}>16$ deserve further attention. Then we list them in Table \ref{tb:50roi} for later studies.
Their positions in the sky are shown in Figure \ref{fig:dis}.
Note that since the radius of the adopted ROI ($2^\circ$) is larger than the interval between two ROI centers, it is possible that several ROIs contain the same excess.
In such a case we only retain the ROI with the largest TS value.

\begin{table*}[!htb]
\caption{Line-like excesses with $\rm TS >16$ identified in this work. }
\begin{ruledtabular}
\begin{tabular}{ccccccccc}
 \#\footnote{The {\tt HEALPix} index in {\tt NESTED} ordering.} &  Energy [GeV] & TS Value\footnote{This column reports the local TS values, the global significances are in fact extremely low due to the very large trial factor. The $TS=24.3$ and $22.4$ correspond to global significances of $0.54\sigma$ and $0.11\sigma$, respectively, while for any other regions the global significances are $<0.1\sigma$.} & TS$_{\rm 2nd}$\footnote{The largest TS values for the second line signals. Please see the main text for details. For the four ROIs with TS$_{\rm 2nd}>4$, we also provide the corresponding energies of the second lines in the brackets (in units of GeV).}  & LON [$^\circ$] & LAT [$^\circ$] & RA [$^\circ$] & Dec [$^\circ$]\\
 \hline
25363 & 74.9 & 24.3 & 0.0  & 182.81 & -15.09 &  74.40 &  18.21 \\
21714 & 17.5 & 22.4 & 0.1  & 114.61 &  -5.98 & 357.97 &  55.93 \\
 9695 & 29.2 & 21.5 & 3.8  & 269.15 &  50.48 & 171.98 &  -6.83 \\
45226 & 16.2 & 21.0 & 0.0  & 272.81 & -78.28 &  20.20 & -37.07 \\
15172 & 48.6 & 20.1 & 6.3(91.3) & 296.32 &  50.48 & 188.56 & -12.17 \\
24414 &  9.4 & 20.0 & 1.2  &  97.73 &  31.39 & 265.83 &  67.66 \\
34394 & 18.2 & 19.6 & 0.9  & 62.58 & -41.81 & 332.47 &   1.38 \\
18272 & 38.4 & 19.6 & 5.8(68.7) & 25.31 &   7.78 & 272.45 &  -3.16 \\
37762 &  5.2 & 19.5 & 1.8  &  125.36 & -58.92 &  14.11 &   3.93 \\
36744 &  5.8 & 19.1 & 3.2  & 37.97 & -12.64 & 296.38 &  -1.35 \\
  278 & 27.0 & 18.8 & 0.6  & 59.77 &  14.48 & 281.32 &  30.20 \\
36061 & 66.6 & 18.6 & 0.0  &  48.52 & -23.32 & 310.58 &   2.24 \\
35642 &  5.6 & 18.6 & 0.8  & 20.39 & -32.80 & 308.78 & -24.17 \\
42727 &  5.4 & 18.5 & 2.3  & 234.84 & -34.95 &  76.89 & -32.24 \\
46981 &  8.7 & 18.3 & 0.0  & 333.98 & -32.80 & 302.38 & -62.61 \\
40543 & 11.8 & 18.3 & 0.8  & 132.19 & -17.58 &  25.25 &  44.40 \\
20889 & 50.6 & 18.1 & 3.6  & 97.73 & -19.47 & 342.51 &  37.39 \\
29481 & 38.4 & 18.0 & 0.8  & 266.48 & -14.48 & 115.85 & -53.82 \\
35865 & 27.0 & 18.0 & 3.9  & 47.11 & -35.69 & 320.62 &  -5.10 \\
10660 &  5.2 & 18.0 & 1.0  &206.72 &  41.01 & 138.95 &  21.83 \\
20103 & 35.5 & 17.9 & 3.5  &344.53 &  17.58 & 239.68 & -29.78 \\
14243 & 10.5 & 17.8 & 0.0  &333.37 &  53.57 & 210.44 &  -5.09 \\
35888 & 26.0 & 17.6 & 9.5(60.4) & 45.00 & -34.95 & 319.13 &  -6.24 \\
 7168 &  9.0 & 17.6 & 2.8  &135.00 &  42.61 & 163.57 &  71.67 \\
14313 & 66.6 & 17.3 & 1.3  &345.37 &  60.43 & 212.36 &   4.16 \\
29638 & 13.3 & 17.2 & 1.6  &270.70 &  -7.18 & 130.02 & -53.51 \\
16786 & 11.8 & 17.2 & 0.2  &  7.73 & -20.74 & 291.84 & -31.05 \\
 7867 &  8.0 & 17.1 & 1.2  & 91.67 &  70.17 & 206.89 &  43.40 \\
28138 & 31.6 & 17.1 & 4.1(64.1) & 186.33 &  24.62 & 115.70 &  33.53 \\
 1666 & 37.0 & 17.0 & 1.3   & 49.92 &  37.17 & 253.34 &  28.86 \\
27865 &  5.4 & 17.0 & 0.0   & 182.11 &  14.48 & 102.60 &  33.75 \\
42296 & 29.2 & 17.0 & 0.0  & 260.08 & -45.78 &  62.38 & -51.43 \\
30431 & 10.5 & 16.9 & 2.3   & 284.06 &   6.58 & 161.66 & -51.66 \\
30245 & 13.3 & 16.8 & 0.5   & 280.55 &  -4.78 & 145.21 & -59.10 \\
19922 & 31.6 & 16.8 & 0.9   & 13.36 &  23.32 & 253.21 &  -5.35 \\
46538 & 35.5 & 16.8 & 3.6   & 346.64 & -38.68 & 311.85 & -51.92 \\
 6201 & 35.5 & 16.8 & 1.8   & 111.80 &  27.28 & 279.44 &  80.10 \\
29823 &  6.1 & 16.8 & 0.5   & 298.12 &  -5.38 & 180.12 & -67.77 \\
38012 & 29.2 & 16.7 & 0.5   & 168.96 & -50.48 &  40.14 &   2.38 \\
24463 &  9.0 & 16.6 & 0.9   & 84.37 &  29.31 & 269.93 &  56.11 \\
28712 & 12.3 & 16.5 & 1.1   & 265.78 & -36.42 &  76.96 & -57.30 \\
22401 & 21.3 & 16.3 & 0.0   & 107.58 &   5.98 & 335.47 &  64.30 \\
 2458 & 10.5 & 16.2 & 0.3   &  28.83 &  41.81 & 243.14 &  14.86 \\
 1240 & 46.8 & 16.2 & 1.5   & 68.91 &  34.95 & 259.22 &  43.61 \\
15976 & 45.0 & 16.1 & 2.4   & 295.91 &  65.70 & 189.97 &   2.98 \\
26070 & 74.9 & 16.1 & 0.0   & 217.27 &   4.78 & 109.09 &  -1.66 \\
23595 & 12.3 & 16.1 & 0.4   & 85.78 &   5.38 & 307.71 &  48.50 \\
20147 &  9.7 & 16.0 & 3.3   & 343.12 &  21.38 & 235.67 & -27.83 \\
15410 &  9.7 & 16.0 & 0.2   & 314.17 &  49.70 & 200.28 & -12.52 \\
27195 & 59.2 & 16.0 & 1.4   & 144.84 &  -1.79 &  52.47 &  54.20 \\
\end{tabular}
\end{ruledtabular}
\label{tb:50roi}
\end{table*}

\end{document}